\documentclass[twocolumn,preprintnumbers,amsmath,amssymb]{revtex4-2}
\usepackage{graphicx}
\usepackage[dvipsnames]{xcolor}
\usepackage{ulem}
\usepackage{url,hyperref}
\usepackage{lipsum}
\newcommand{\be}{\begin{equation}}
\newcommand{\ee}{\end{equation}}
\newcommand{\beq}{\begin{equation}}
\newcommand{\eeq}{\end{equation}}
\newcommand{\bea}{\begin{eqnarray}}
\newcommand{\eea}{\end{eqnarray}}
\newcommand{\eml}{\end{mathletters}}

\newcommand{\hf}{|\text{HF}\rangle}

\begin{document}

\title{Reduced basis emulation of pairing in finite systems}

\author{V.V. Baran$^{1,2,3}$}
\email[]{virgil.v.baran@unibuc.ro}
\author{D. R. Nichita$^{2,3}$}

\affiliation{
$^1$Research Institute of the University of Bucharest (ICUB), 050107 Bucharest, Romania\\
$^2$Faculty of Physics, University of Bucharest, 405 Atomi\c stilor, Bucharest-M\u agurele, RO-077125, Romania\\
$^3$"Horia Hulubei" National Institute of Physics and 
Nuclear Engineering, 30 Reactorului, RO-077125, Bucharest-M\u agurele, Romania
}

\begin{abstract}
\begin{description}
\item[Background] In recent years, reduced basis methods (RBMs) have been adapted to the many-body eigenvalue problem and they have been used, largely in nuclear physics, as fast emulators able to bypass expensive direct computations while still providing highly accurate results. 

\item[Purpose] This work is meant to show that the RBM is an efficient and accurate emulator for the strong correlations induced by the pairing interaction in a variety of finite systems like ultrasmall superconducting grains, interacting topological superfluids and mesoscopic hybrid superconductor-semiconductor devices, all of which require an expensive, beyond-mean-field, particle-number conserving description. 
\item[Method] These systems are modelled by the number-conserving Richardson pairing Hamiltonian and its appropriate generalizations. Their ground state is solved for exactly using the Density Matrix Renormalization Group (DMRG). The reduced basis is assembled iteratively from a small number of exact ground state vectors, well-chosen from across the relevant parameter space using a fast estimate of the emulation error and a greedy
local optimization algorithm.
\item[Results]  The reduced basis emulation is found to accurately describe the weak-to-strong pairing cross-over in small grains, the third-order topological phase transition of the interacting Richardson-Kitaev chain, and the complex charge stability diagram of a hybrid quantum dot - superconductor device. For all considered systems the emulation error decreases exponentially with the  reduced basis dimensionality. The number of basis vectors necessary to reach a fixed emulation accuracy only shows a modest linear growth with increasing system size.
\item[Conclusions] RBMs are  confirmed to be cheap and accurate emulators for the widely encountered superconducting phenomena. Capable of providing orders of magnitude computational speed-up with respect to approaches based only on traditional many-body solvers, they open new possibilities in building and solving models of interacting many-body systems and in better interfacing them with experimental design and data analysis.

\end{description}
\end{abstract}

\maketitle

\section{Introduction}
Strongly correlated  systems are ubiquitous in several fields of fundamental physics including condensed matter \cite{vonDelft2001Apr, Dukelsky2004Aug} and nuclear physics \cite{Broglia2012}. Across these fields, the quantum-mechanical treatment of many interacting particles is extremely challenging both formally and computationally, which has led to an increased interest in building efficient emulators whose predictions go beyond what is possible with direct calculations and make many-query computations more affordable \cite{Carleo2019Dec,Thiagarajan2020Nov,Kasim2021Dec}.

Recently, the Eigenvector Continuation (EC) approach was introduced in Ref. \cite{frame2018} as an emulator capable of efficient interpolations and extrapolations for the extremal eigenstates of a Hamiltonian defined by one or more variable parameters. Nuclear physicists have been using it for the emulation of scattering and reactions \cite{furnstahl2020,melendez2021,drischler2021,bai2021,bai2022,zhang2022,yapa2022} and shell model calculations \cite{yoshida2022}, as a perturbation theory re-summation tool \cite{demol2020,demol2021,companys2022}, and for sensitivity analysis and uncertainty quantification \cite{ekstrom2019,konig2020,djarv2022}. Furthermore, it has been shown to provide a universal set of states capable of reproducing the entire single-particle spectrum across the nuclear chart with remarkable accuracy \cite{anderson2022}.

The success of EC relies on the smooth eigenstate behaviour over the parameter manifold being effectively limited to a very low-dimensional subspace (compared to the full Hilbert space dimensionality). In practice, this means that the EC ansatz may be computed as a linear combination of (a small number of) ``exact" solutions of the considered problem, by solving a generalized eigenvalue problem in the restricted subspace. The ``exact" or ``truth" direct solutions are to be obtained using any high-fidelity (albeit usually computationally expensive) numerical many-body method.

More recently, EC has naturally been recognized \cite{bonilla2022,Melendez2022Sep} as being an instance of a larger class of reduced-basis methods (RBMs) \cite{Almroth2012May,Quarteroni,Hesthaven}, part of the general framework of reduced-order models \cite{ROM, Brunton2019Feb,Melendez2022Sep}. Simultaneously, the idea of using RBMs in the context of eigenvalue problems has been independently rediscovered and applied to build surrogate models (emulators) for the efficient determination of phase diagrams in quantum spin models in Ref. \cite{rizzi2022}.

Given that RBMs have only been used in recent years for the study of quantum many-body systems, their applicability remains to be tested in various scenarios. It is the purpose of this work to show that the strong correlations induced by the pairing interaction in the ground state of finite systems may be successfully emulated with RBMs.

In Section \ref{s2}, we shortly review the reduced-basis methodology in the context of eigenvalue problems. Section \ref{s3} is devoted to the RBM treatment of the generic pairing Hamiltonian, also known as the Richardson Hamiltonian. It used to describe pairing between protons or between neutrons in  atomic nuclei \cite{Broglia2012} and electron pairing in ultrasmall superconducting grains \cite{vonDelft2001Apr, Dukelsky1999Jul} or small superconducting islands \cite{Pavesic2021, Pavesic2022, Saldana2022Feb,Saldana2022Apr, Malinowski2022Oct} in condensed matter physics. In Section \ref{s4} we obtain the first RBM description of a topological phase transition \cite{Ortiz2014Dec}, and  in Section \ref{s5} we show that the RBM is able to successfully emulate the complex charging patterns of an interacting quantum dot in contact with a superconducting island \cite{Pavesic2021}. Finally, in Section \ref{s6} we draw conclusions.

\section{Reduced-basis methodology}
\label{s2}

We present here a short review of the general reduced-basis methodology employed in this work, for self-consistency. For a more detailed treatment, we refer the reader to the Refs. \cite{Melendez2022Sep, rizzi2022}.

All systems under consideration are modelled by Hamiltonians that may be written in a so-called affine decomposition (see also the specific examples in the next sections),
\beq
\label{aff}
\mathcal{H}({\boldsymbol{\xi}})=\sum_{p=0}^{N_\mathcal{H}} f_p(\boldsymbol{\xi}) H_p~,
\eeq
where $\boldsymbol{\xi}=(\xi_1,\xi_2,\dots,\xi_n)$ is a set of $n$ control parameters in the domain of interest. Here, $f_p(\boldsymbol{\xi})$ are a set of $N_\mathcal{H}+1$ functions dictating the full parameter dependence of the system's Hamiltonian $\mathcal{H}({\boldsymbol{\xi}})$, with the corresponding $H_p$'s being parameter-independent. By convention, we denote the parameter-independent part of $\mathcal{H}({\boldsymbol{\xi}})$ by $H_0$, and thus we take $f_0(\boldsymbol{\xi})=1$.  

The core RBM idea is to construct a low-dimensional effective representation $\psi^{(\text{rb})}({\boldsymbol{\xi}})$ for the ground state of $\mathcal{H}({\boldsymbol{\xi}})$ that is sufficiently accurate over the parameter domain of interest. In doing so, we only employ a relatively small number $N_{\text{rb}}$ of sample or training points ${\boldsymbol{\xi}}_k$  where ``exact" solutions $\psi^{(ex)}_k$ of the Hamiltonian problem are actually computed, i.e.
\beq
|\psi^{(\text{rb})}({\boldsymbol{\xi}})\rangle=\sum_{k=1}^{N_{\text{rb}}} c_k({\boldsymbol{\xi}})|\psi^{(ex)}_k\rangle~,
\eeq
with $N_{\text{rb}}$ being much smaller than the dimension of the full many-body Hilbert space.
Here, the expansion coefficients $c_k({\boldsymbol{\xi}})$ for any parameter values may be obtained upon solving the effective generalized eigenvalue problem for $h_{k\ell}({\boldsymbol{\xi}})=\langle\psi^{(\text{ex})}_k|\mathcal{H}({\boldsymbol{\xi}})|\psi^{(\text{ex})}_\ell\rangle$ in the (non-orthogonal) reduced basis of $\psi^{(\text{ex})}$'s,
\beq
\label{gep}
h\, c=E^{(\text{rb})}\, S\, c~,
\eeq
involving the overlap matrix 
\beq 
\label{overlap}
S_{k\ell}\equiv\langle\psi^{(ex)}_k|\psi^{(ex)}_\ell\rangle~.
\eeq

Regarding the choice of sampling points, one efficient way to sample the parameter space is through an active learning protocol which combines a fast estimate of the emulation error and a greedy optimization algorithm that becomes progressively more
accurate \cite{sarkar2022,rizzi2022}. In this so-called ``offline" emulation phase, the reduced basis is constructed iteratively by repeating two main steps. 

The first step is an evaluation over the relevant parameter space of the Residual
    \beq
    \label{res}
    \text{Res}({\boldsymbol{\xi}})\equiv
\left\lVert\mathcal{H}({\boldsymbol{\xi}})|\psi^{(\text{rb})}({\boldsymbol{\xi}})\rangle-E^{(\text{rb})}({\boldsymbol{\xi}})|\psi^{(\text{rb})}({\boldsymbol{\xi}})\rangle\right\rVert~,
    \eeq
which quantifies the degree to which the exact Schr\"odinger equation is satisfied by the current reduced-basis approximate solution $\psi^{(\text{rb})}$. Note that only the emulated information is used in this step and there is no need to perform the error evaluation using new costly ``exact" solutions  (which would defeat the purpose of the emulation). 

The location in the parameter domain of the maximum residual indicates where the emulation is least accurate and would benefit the most from a new sampling point (see also Appendix \ref{app} for more details on this point). It is at the residual's maximum location that, in the second step, we run the ``exact" solver and add the resulting vector to the reduced basis, thus increasing $N_{\text{rb}}$ by one unit.

These steps are repeated until the emulation error (as quantified by the above defined Residual) decreases below a suitably chosen threshold. On one hand, the success of the RBM in  reaching rapidly a good enough emulation accuracy within a low-dimensional subspace is naturally related to the high linear dependence of the basis of training vectors $\psi^{(\text{ex})}_k$ \cite{Quarteroni,bonilla2022}. 

On the other hand, this may lead to a numerically singular overlap matrix $S$ and thus to numerical instabilities in the solution to Eq. (\ref{gep}). It is customary to improve upon this by orthogonalizing the $\psi^{(\text{ex})}_k$'s, leading to a standard eigenvalue problem. In practice, we perform this orthogonalization by solving the eigenvalue problem for the Hermitian overlap matrix $S \widetilde{c}_a = s_a \widetilde{c}_a$ and using its orthogonal eigenvectors $\widetilde{c}_a$ to build a new and numerically stable reduced basis. 

Within this proper orthogonal decomposition (POD) approach \cite{Quarteroni,bonilla2022,rizzi2022} we have the option of further compressing our basis by discarding those $\widetilde{c}_a$ vectors  corresponding to eigenvalues $s_a$ lower than a certain tolerance value, thus capturing an effective low-dimensional representation of the full set of “exact”
ground-state solutions $\psi^{(ex)}_k$. The largest-$s$ $\widetilde{c}$-vector can be seen as the `average' ground state, while the small-$s$ $\widetilde{c}$-subspace accommodates fine ground state variations over the sampled parameter domain. That the small-$s$ contributions are mostly decoupled from the low-energy subspace may be seen also from the equivalent isospectral form of Eq. (\ref{gep}) involving $S^{-1/2}hS^{-1/2}$.
    We quantitatively discuss the effects of this lossy basis compression on the emulation performance in the next section (see e.g. Fig. \ref{fig5}).

The reduced basis assembled during the emulator's training in the offline phase is guaranteed to approximate sufficiently well our system's ``exact" ground state over the entire relevant parameter domain. This basis is then used in the final ``online" emulation phase to evaluate efficiently the behaviour of any affinely decomposable observable of interest  $\mathcal{A}({\boldsymbol{\xi}})=\sum_{p=1}^{N_\mathcal{A}}g_p(\boldsymbol{\xi}) A_p$ using the reduced-basis representation
\beq
\begin{aligned}
A^{(\text{rb})}({\boldsymbol{\xi}})&=\langle\psi^{(\text{rb})}({\boldsymbol{\xi}})|\mathcal{A}({\boldsymbol{\xi}})|\psi^{(\text{rb})}({\boldsymbol{\xi}})\rangle\\
&=\sum_{p=1}^{N_\mathcal{A}}\sum_{k,\ell=1}^{N_{\text{rb}}}g_p(\boldsymbol{\xi}) c_k^*({\boldsymbol{\xi}})c_\ell({\boldsymbol{\xi}})\langle \psi^{(ex)}_k|A_p|\psi^{(ex)}_\ell\rangle,
\end{aligned}
\eeq
where the parameter-independent matrix elements of the various $A_p$'s in the $\psi^{(ex)}_k$ basis need only be evaluated once using ``exact" methods. If employing the above POD approach, any operator matrix element in  the compressed basis may be trivially computed from the matrix elements in the original $\psi^{(\text{ex})}$ basis.  In the case of observables with non-affine parameter dependencies, techniques such as the Empirical Interpolation Method may be used to construct an affine approximation \cite{Barrault2004Nov,Grepl2007May}.

We finally note that the RBM framework is blind to the actual  method employed for obtaining the ``exact" ground state solutions: the only input needed for a minimal RBM emulator consists of the matrix elements $\langle\psi^{(\text{ex})}_k|H_p|\psi^{(\text{ex})}_\ell\rangle$ of the Hamiltonian affine components $H_p$ together with the overlaps  $\langle\psi^{(\text{ex})}_k|\psi^{(\text{ex})}_\ell\rangle$ used to construct the effective eigenvalue problem of Eq. (\ref{gep}). Additionally, the matrix elements $\langle\psi^{(\text{ex})}_k|H_p H_{p'}|\psi^{(\text{ex})}_\ell\rangle$ must be provided to the RBM emulator for evaluating the Residual in Eq. (\ref{res}) during the offline phase, and the matrix elements of any operator of interest $\langle \psi^{(ex)}_k|A_p|\psi^{(ex)}_\ell\rangle$ for the online phase.

All ``exact" solutions are obtained here by the Density Matrix Renormalization Group (DMRG) technique \cite{White1992Nov} in the Matrix Product State (MPS) formulation  \cite{Schollwock2011Jan}, using the Itensor library \cite{itensor,itensor-r0.3} for the numerical implementation. The latter allows for complete control over the numerical accuracy of the obtained solutions, while providing MPS tools for evaluating efficiently the matrix elements mentioned in the previous paragraph. We employed a $10^{-9}$ energy convergence tolerance for the DMRG sweeps in the presence of a maximum bond dimension of $2000$ and utilized the ITensor `noise' term to ensure that the energy global minimum is reached.

\section{Emulating the Richardson Pairing Hamiltonian}
\label{s3}

\subsection{Exact ground state structure}
We first discuss qualitatively the structure of the ground state for the particle-number conserving generic $s$-wave singlet-pairing Hamiltonian of spin-1/2 fermions
\beq
\label{ham1}
\begin{aligned}
H(G)&\equiv\sum_{i=1}^L\sum_{\sigma=\uparrow,\downarrow} \epsilon_i\, c^\dagger_{i,\sigma} c_{i,\sigma}  - G \sum_{i,j=1}^L c^\dagger_{i,\uparrow}c^\dagger_{i,\downarrow}c_{j,\downarrow} c_{j,\uparrow}\\
&\equiv\sum_{i=1}^L \epsilon_i\, N_i - G \sum_{i,j=1}^L P^\dagger_{i}P_{j}~,
\end{aligned}
\eeq
where $i$ indicates one of the $L$ pairs of conjugated degenerate single particle levels with energy $\epsilon_{i}=\epsilon_{i,\uparrow}=\epsilon_{i,\downarrow}$ and $G$ denotes the pairing interaction strength. In the following, we will take the single particle levels to be uniformly distributed with energy spacing $\epsilon=\epsilon_{i+1}-\epsilon_{i}$.

For simplicity, we restrict ourselves to even particle-number systems with no unpaired particles. Then, in the absence of the interaction term the ground state is given by the Hartree-Fock product state
\beq
\label{hf}
\hf=\prod_{i=1}^{N_p} P^\dagger_{i} \,|0\rangle\qquad (G=0)~,
\eeq
where $N_p$ is the number of pairs in the system. The weak pairing regime $G\ll \epsilon$ may be described to a very good approximation by coherent pair-excitations through a Coupled-Cluster Doubles variational ansatz \cite{ccdpbcs}
\beq
|\psi\rangle\simeq \exp\left[\sum_{ij}z_{ij} P^\dagger_i P_j\right] \hf  \qquad (G\ll \epsilon)~.
\eeq

The opposite limit of very strong pairing with $G\ggg \epsilon$ or equivalently $\epsilon=0$ is known as the Seniority Model \cite{Ring} in nuclear physics and as the Zero Bandwidth Model \cite{Zitko2022Jul} in the condensed matter physics community, and has a product of identical collective pairs as the exact ground state,
\beq
|\psi\rangle=\left(\sum_{i=1}^{L}P^\dagger_i\right)^{N_p}|0\rangle\qquad (G\ggg \epsilon)~.
\eeq

The cross-over regime between weak and strong pairing (where the exact ground state structure changes qualitatively from coherent pair-excitations to a collective pair condensate) is situated on an energy scale comparable to the critical BCS pairing strength $G_\text{cr}$ \cite{ccdpbcs}. At this value of the pairing strength, the $U(1)$-particle-number-breaking BCS solution
collapses and gives zero pairing correlations. This is
contrary to the expectation from infinite systems where $G_{\text{cr}}$ vanishes and the
pairing correlations depend exponentially on the pairing strength at weak coupling. For the range of system sizes considered in this work, the critical pairing strength varies from $G_{\text{cr}}\simeq 0.32\epsilon$ for $L=2N_p=12$ to $G_{\text{cr}}\simeq 0.16\epsilon$ for $L=2N_p=200$.

Note that while the previous considerations apply also to more general pairing Hamiltonians involving non-constant matrix elements of the pairing potential, the specific Hamiltonian of Eq. (\ref{ham1}) actually admits an exact solution originally discovered by Richardson \cite{Richardson1966Jan}. The Richardson solution provides a unified description of all pairing regimes in terms of a product of distinct collective pair operators
\beq
\label{richardson}
|\psi\rangle=\prod_{\alpha=1}^{N_p}\left(\sum_{i=1}^{L}\frac{P^\dagger_i}{2\epsilon_i-e_\alpha} \right)|0\rangle \qquad (\text{any } G)~,
\eeq
each collective pair being defined by a different pair energy $e_m$. The pair energies may be found for any value of the pairing strength $G$ by solving a set of coupled nonlinear equations \cite{Dukelsky2004Aug}.

\subsection{RBM results}

\begin{figure}[ht!]
\centering
\includegraphics[width=0.5\textwidth]{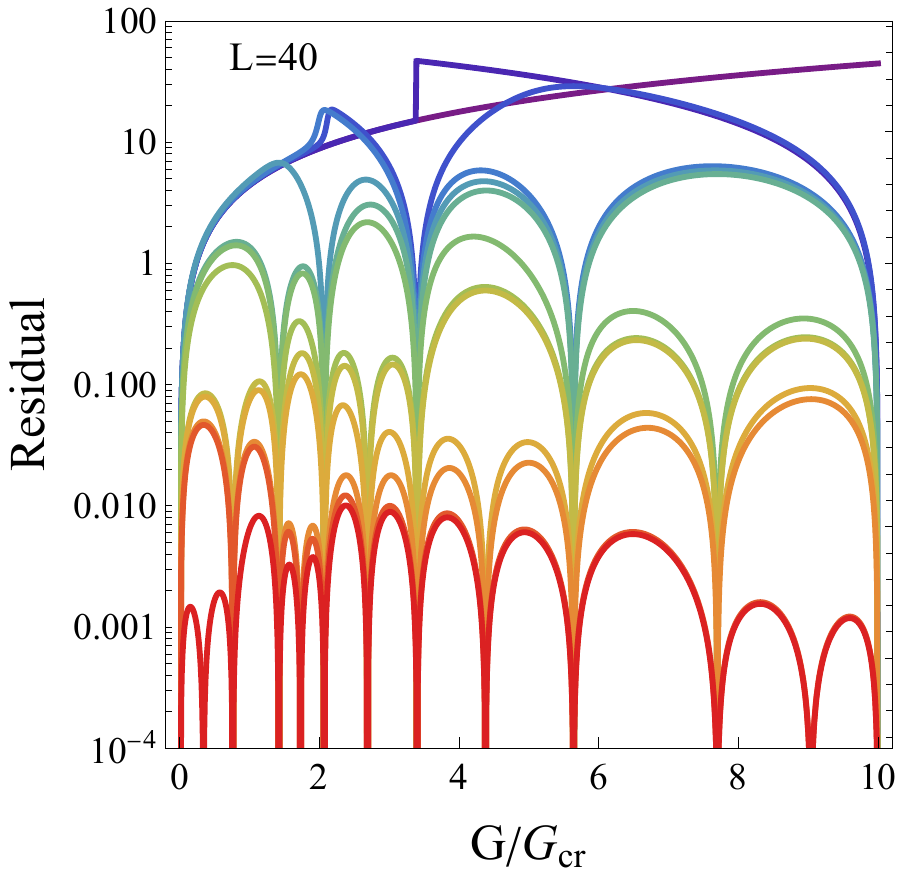}
\caption{Evolution of the Residual (\ref{res}), in units of the level spacing $\epsilon$, during the offline emulation phase of the pairing Hamiltonian (\ref{ham1}) for a system of $N_p=20$ pairs distributed over $L=40$ levels. The color spectrum (violet, blue, green, orange, red) correlates with the increasing iteration number of the greedy self-learning algorithm. }
\label{fig1}
\end{figure}

  We are now in a position to benchmark the RBM capabilities in emulating the weak to strong pairing cross-over. For this purpose, we consider the Richardson pairing Hamiltonian of Eq. (\ref{ham1}) with the pairing strength $G$ as the control parameter, $\xi=G$, and we take the level spacing $\epsilon=1$ as the unit of energy. We sample the interval $0\leq G\leq 10G_{\text{cr}}$ using the greedy algorithm described in the previous section, leading to the Residual evolution over 14 iterations illustrated in Fig. (\ref{fig1}) for $N_p=20$ pairs distributed on $L=40$ levels (in this section we consider all systems at half-filling, $L=2N_p$). 
  
  We take the first sampling point at $G=0$ and thus $|\psi^{\text{(ex)}}_1\rangle=\hf$ state as the initial reduced-basis vector. In this first iteration the emulation accuracy naturally degrades when moving away from the weak pairing regime. The Residual thus reaches its maximum at the other end of the interval ($G=10G_{\text{}cr}$), where the next basis vector $\psi^{\text{(ex)}}_2$ is computed by a DMRG evaluation.
  
  In the presence of sampling points situated at both ends of our parameter interval the new Residual maximum is now more centrally located around $G=3.4G_{\text{cr}}$, where we compute with DRMG a new solution $\psi^{\text{(ex)}}_3$. This procedure is repeated until achieving the desired  emulation accuracy, as given by the global maximum of the Residual.

  \begin{figure}[ht!]
\centering
\includegraphics[width=0.5\textwidth]{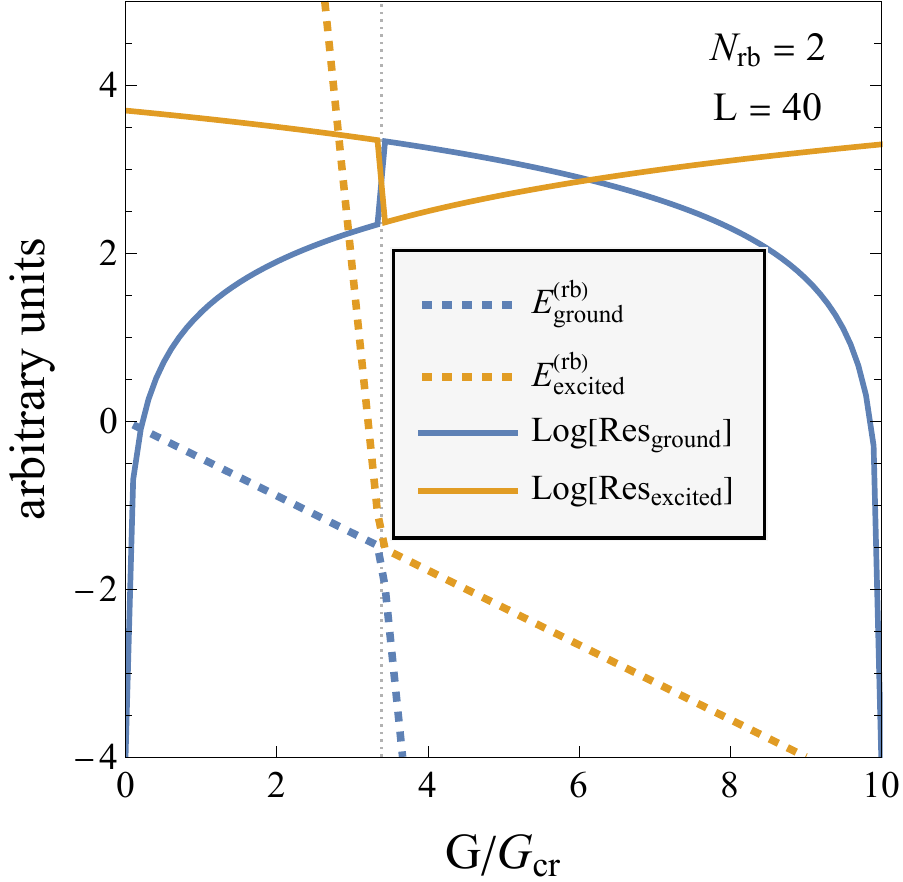}
\caption{Energies $E^{\text{rb}}$ and Residuals (\ref{res}) corresponding the ground and excited states in a two-dimensional reduced subspace with sampling vectors situated at the interval's ends.}
\label{fig2}
\end{figure}

  As also remarked in Ref. \cite{rizzi2022} and contrary to a naive expectation, the decrease of the Residual's maximum from one iteration of the greedy sampling algorithm to the next is not always monotonic. Additionally, more or less pronounced  discontinuities are present in the Residual profile in Fig. (\ref{fig1}) during the early iterations.

  To better understand this behaviour, we give in Fig. (\ref{fig2}) more details on the second iterative step $N_{\text{rb}}=2$ involving sampling points at both interval endpoints.
  Fig. (\ref{fig2}) shows how the $\psi^{(\text{ex})}_1$-dominated vector and the $\psi^{(\text{ex})}_2$-dominated vector (having continuous energies and Residuals  over the entire parameter range) exchange roles as ground and excited states around $G=3.4G_{\text{cr}}$. At this point, the $\psi^{(\text{ex})}_1$-dominated ground state suddenly becomes $\psi^{(\text{ex})}_2$-dominated and experiences a Residual jump. For denser $N_{\text{rb}}>2$ samplings, the Residual mismatch for the various $\psi^{(\text{ex})}_k$-dominated vectors is naturally lower. As a consequence the jumps in the ground state Residual become less pronounced, as also seen in Fig. (\ref{fig1}).

\begin{figure}[hb!]
\centering
\includegraphics[width=0.5\textwidth]{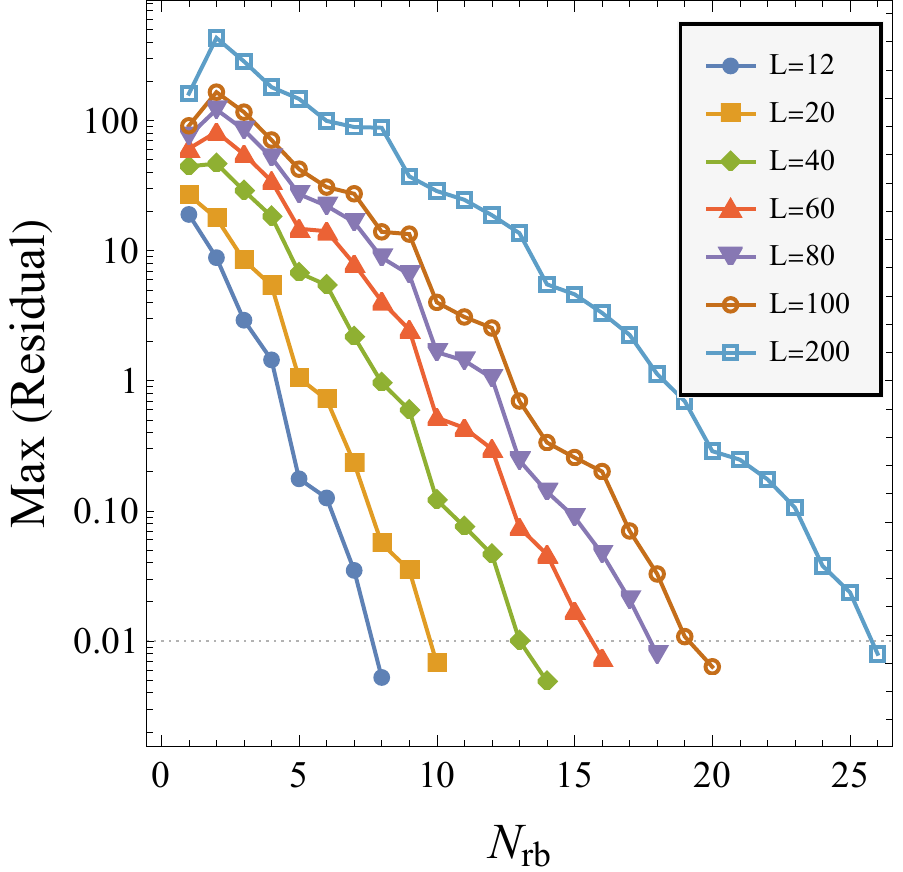}
\caption{Evolution of the Residual's maximum, in units of the level spacing $\epsilon$, versus the reduced basis dimension $N_{\text{rb}}$ during the offline emulation phase of the pairing Hamiltonian (\ref{ham1}) for various systems sizes, at half filling $L=N_p$.}
\label{fig3}
\end{figure}

 The Residual jump of the ground state at $N_{\text{rb}}=2$ is amplified when increasing the system's size, as shown in Fig. (\ref{fig3}). Nevertheless, in all the tested examples, ranging from small $L=12$ to large $L=200$ systems, the Residual maximum decays exponentially beyond $N_{\text{rb}}=2$. This leads to an approximately linear increase with the system's size $L$ of the number of basis states $N_{\text{rb}}$ needed to reach our chosen threshold for the emulation error of Res$/\epsilon=10^{-2}$, of about one additional state per 10 extra levels. Note also that we choose to scale up the system while keeping intact the sampling interval $0\leq G\leq 10G_{\text{cr}}$ in order to emulate the weak to strong pairing cross-over under controlled conditions. The reduction of the critical pairing strength $G_{\text{cr}}$ by a factor of two from $L=12$ ($G_{\text{cr}}\simeq 0.32\epsilon$) to $L=200$ ($G_{\text{cr}}\simeq 0.16\epsilon$) would then imply at most a doubling of the reduced-basis growth rate if keeping the absolute $G/\epsilon$ interval constant during system scaling.

\begin{figure}[ht!]
\centering
\includegraphics[width=0.5\textwidth]{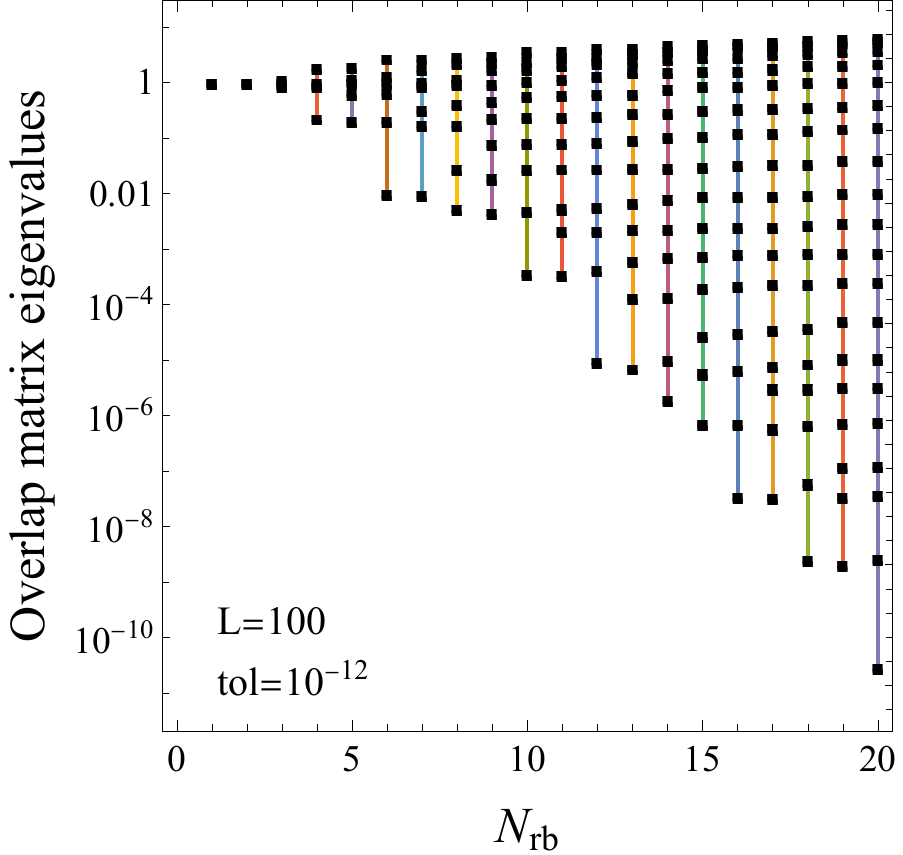}
\caption{Evolution of the overlap matrix (\ref{overlap}) eigenvalues with increasing reduced basis dimension $N_{\text{rb}}$ during the offline emulation phase of the pairing Hamiltonian (\ref{ham1}).}
\label{fig4}
\end{figure}

 The exponentially fast decay of the residual during the greedy sampling algorithm is related to high linear dependence of the basis vectors (see the discussion in Section \ref{s2}). This in turn implies an exponential suppression of new overlap-matrix eigenvalues during the reduced basis construction (offline) phase, which is confirmed in Fig. (\ref{fig4}). 

 \begin{figure}[ht!]
\centering
\includegraphics[width=0.5\textwidth]{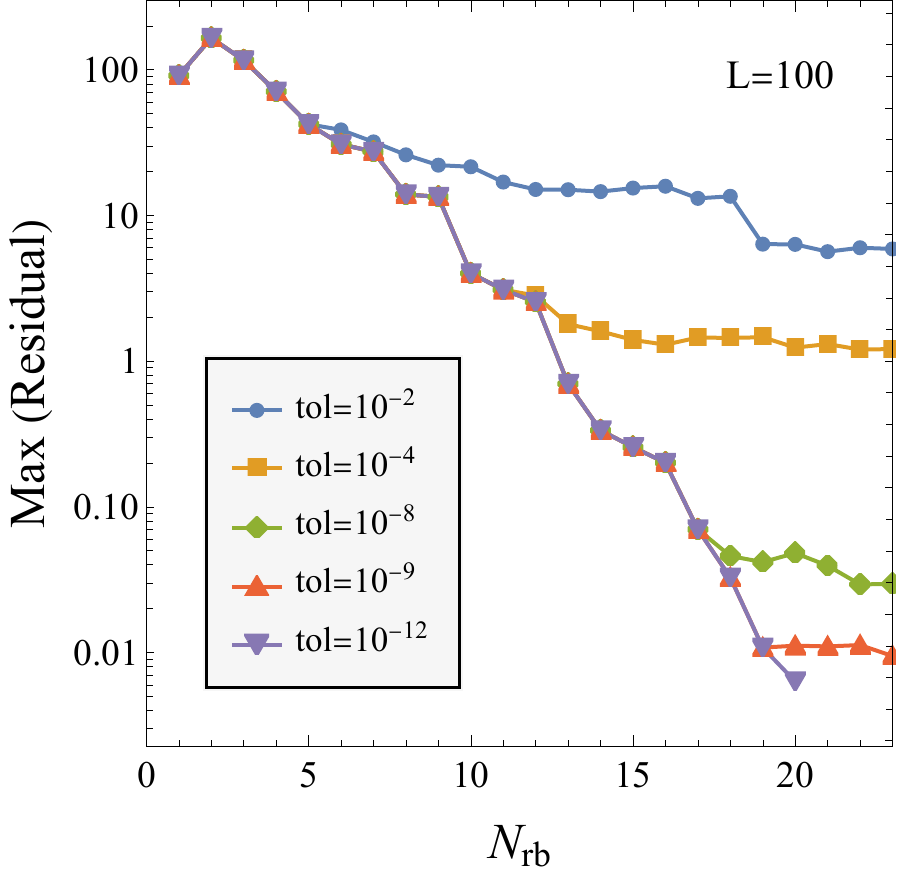}
\caption{Evolution of the Residual's maximum, in units of the level spacing $\epsilon$, versus the reduced basis dimension $N_{\text{rb}}$ during the offline emulation phase of the pairing Hamiltonian (\ref{ham1}), for various POD compression tolerances.}
\label{fig5}
\end{figure} 
 
 We note however that the POD basis compression losses may have a large impact on the emulator's ability to actually reach the target accuracy. We explore this in Fig. (\ref{fig5}) where the evolution of the Residual's maximum is shown for various tolerances of the overlap matrix eigenvalues. If the tolerance is not low enough (and too many vectors are discarded from the reduced basis) the remaining subspace may not be large enough to accommodate all relevant ground state variations over the parameter domain. This leads to a saturation in the Residual's decay beyond a basis dimensionality for which any new overlap-matrix eigenvalue will not exceed the tolerance value. Except for Fig. (\ref{fig5}), all other results shown in this section are obtained without any POD basis compression.

   \begin{figure}[ht!]
\centering
\includegraphics[width=0.5\textwidth]{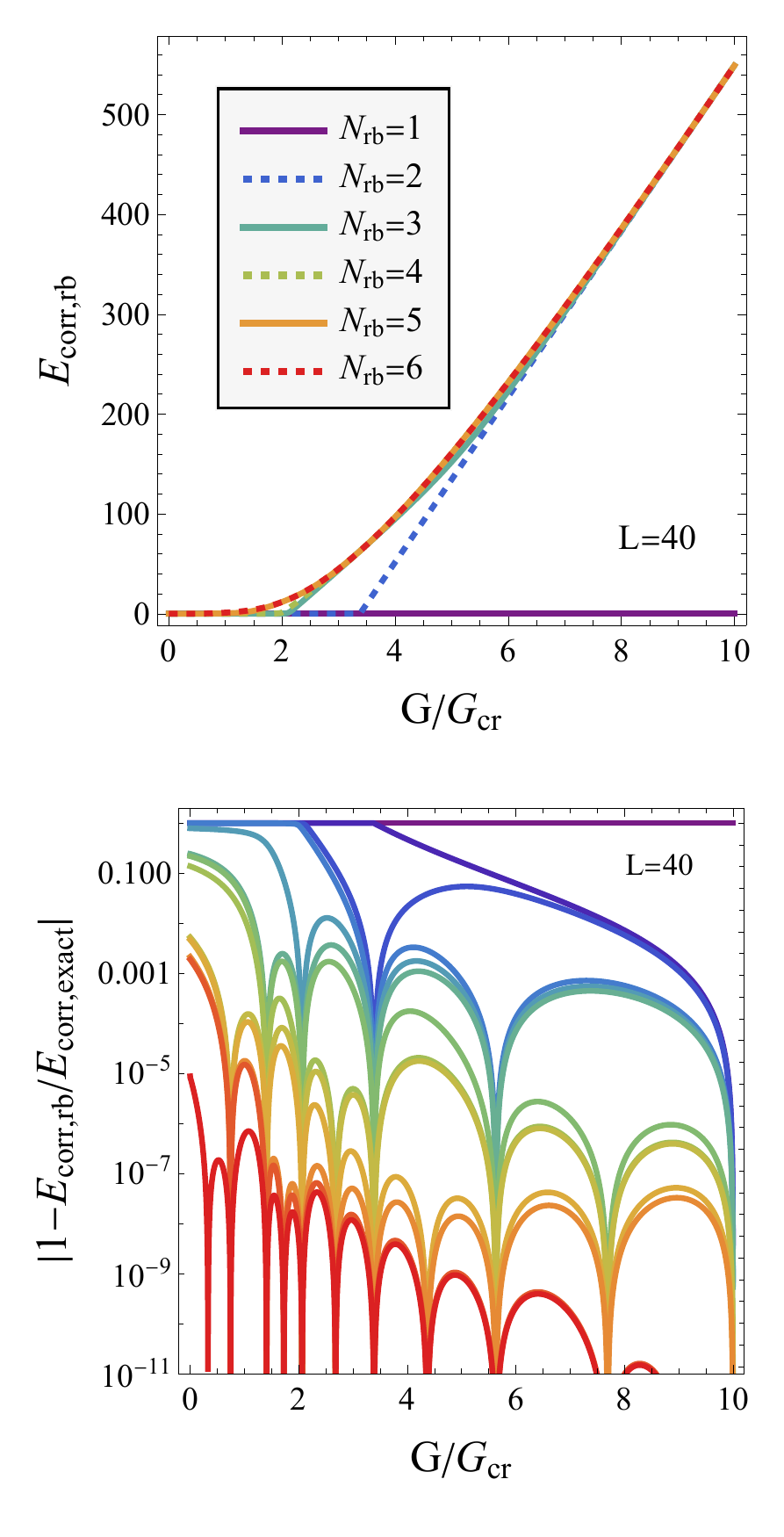}
\caption{Evolution of the reduced basis correlation energy $E_\text{corr}$ (\ref{ecorr}) (top panel), in units of the level spacing $\epsilon$, and the corresponding error relative to its exact value (bottom panel) during the offline emulation phase.}
\label{fig6}
\end{figure}

 \begin{figure}[ht!]
\centering
\includegraphics[width=0.5\textwidth]{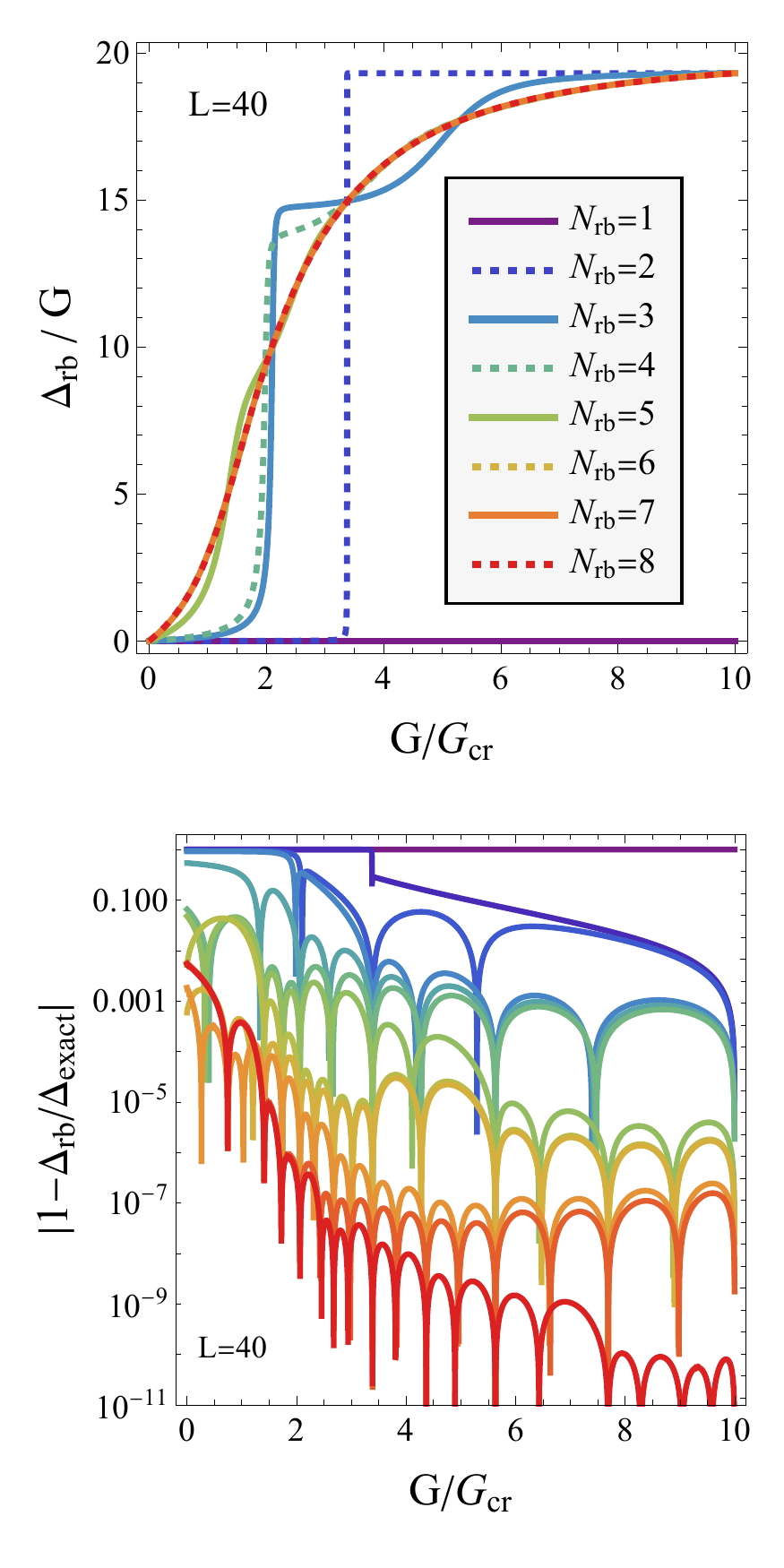}
\caption{Evolution of the reduced basis canonical gap $\Delta$ (\ref{gap}) (top panel), and the corresponding error relative to its exact value (bottom panel), during the offline emulation phase.}
\label{fig7}
\end{figure}

We conclude this section by assessing the quality of the greedy algorithm in converging to the exact solution of the Richardson pairing Hamiltonian. As a first convergence metric, we use the correlation energy
\begin{equation}
\label{ecorr}
    E_{\text{corr}}(G)\equiv  E_{\text{HF}}(G)-E_{\text{ground}}(G)~,
\end{equation}
 where $E_{\text{HF}}(G)=\langle \text{HF}|H(G)|\text{HF}\rangle=\sum_{i=1}^{N_p}(2\epsilon_i-G)$ is the energy of the Hartree-Fock state. $E_{\text{corr}}$ measures the amount of correlations captured by the wavefunction under consideration beyond the Hartree-Fock state and is thus  sensitive on the wavefunction's structure especially in the weak pairing regime, as can be seen in Fig. (\ref{fig6}b). Even with the $|\text{HF}\rangle$ state as one of the sampling points, in the extremely weak pairing regime $G\simeq 0$ the error in the correlation energy (relative to its exact value) remains finite due to the vanishing of $E_{\text{corr}}(G=0)$. This gives an indication on the rate at which the Hartree-Fock state is approached by our approximate solution, which turns out to be similar to that obtained using a more complex variational ansatz \cite{ccdpbcs} (for the chosen Residual accuracy goal of $10^{-2}\epsilon$).
 
 As a second convergence metric, we consider the canonical gap 
 \begin{equation}
 \label{gap}
\Delta=G \sum_{i=1}^L \sqrt{n_i\left(1-n_i\right)}
\end{equation}
where $n_i = \langle N_i\rangle/2$  indicates the occupation probability
of each level $i$. Due to its dependence on the occupation
probabilities, this quantity exhibits a more pronounced
sensitivity to the structure of the wave function than the
correlation energy. We show in Fig. (\ref{fig7}a) the profile of $\Delta/G$ obtained during the first iterations of the greedy algorithm, which converges rapidly towards the exact curve in an oscillatory manner. This leads to an increased number of zeros in the gap error presented in  (\ref{fig7}b), corresponding both to  the sampling points and to the intersections of the oscillating approximate profile with the exact curve. Similarly to the case of the correlation energy, the smallness of the gap in the weak pairing regime leads to the errors being largest there. However, the oscillatory behaviour of the reduced-basis gap approximation  leads to a non-monotonic error decrease at  $G\simeq 0$. 

The success of the RBM emulation of the generic Richardson pairing Hamiltonian raises the question of whether RBMs can provide benefits in modelling more complex condensed matter systems like topological superfluids or hybrid superconductor-semiconductor nano-devices, which we investigate in the following.

\section{Emulating Finite Topological Superfluids}
\label{s4}
\subsection{The Richardson-Gaudin-Kitaev chain}

As a first generalization of the Richardson pairing Hamiltonian, we consider in this section the Richardson-Gaudin-Kitaev (RGK) chain introduced in Ref. \cite{Ortiz2014Dec} as a key example of an interacting, particle-conserving, fermionic superfluid in one
spatial dimension displaying a topologically non-trivial superfluid phase. We note that, in the context of topologically protected quantum computation, determining the consequences of particle-number conservation on the properties of Majorana modes and on their braiding statistics (usually discussed within a mean-field treatment) has attracted a lot of attention in recent years and is  an open problem at the time of this writing, see Ref. \cite{Lin2022Dec} and references therein.

Being formulated within the number-conserving setting, the one-dimensional RGK Hamiltonian of Ref. \cite{Ortiz2014Dec} includes
long-range interactions in order to evade the Mermin-Wagner-Hohenberg (or Coleman in field theory) theorem and to display a true
gap in the thermodynamic limit. Its explicit form in the momentum representation is given by
\begin{equation}
\label{RKG}
H_{\mathrm{RGK}}=\sum_{k \in \mathcal{S}_k^\phi} \varepsilon_k {c}_k^{\dagger} {c}_k-8 G \sum_{k, k^{\prime} \in \mathcal{S}_{k+}^\phi} \eta_k \eta_{k^{\prime}} {c}_k^{\dagger} {c}_{-k}^{\dagger} {c}_{-k^{\prime}} {c}_{k^{\prime}}
\end{equation}
for spinless fermions $c^\dagger_k$, with momentum-dependent single-particle spectrum
\begin{equation}
\varepsilon_k=-2 t_1 \cos k-2 t_2 \cos 2 k,
\end{equation}
where $t_1$ and $t_2$ are the nearest and next-nearest neighbor hopping amplitudes, and  a momentum-dependent modulation of the interaction strength
\begin{equation}
\eta_k=\sin (k / 2) \sqrt{t_1+4 t_2 \cos ^2(k / 2)}
\end{equation}
which is odd in $k$, as is characteristic of $p$-wave superconductivity. The specific values of the allowed momenta in the cases of interest below, i.e. for periodic ($\phi=0$) and antiperiodic ($\phi=2\pi$) boundary conditions on a chain of $L$ sites, are $\mathcal{S}_k^0=\mathcal{S}_{k+}^0 \oplus \mathcal{S}_{k-}^0 \oplus\{0,-\pi\}$  and $\mathcal{S}_k^{2 \pi}=\mathcal{S}_{k+}^{2 \pi} \oplus \mathcal{S}_{k-}^{2 \pi}$, with $\mathcal{S}_{k \pm}^0=L^{-1}\{\pm 2 \pi, \pm 4 \pi, \cdots, \pm(\pi L-2 \pi)\}$ and $\mathcal{S}_{k \pm}^{2 \pi}=L^{-1}\{\pm \pi, \pm 3 \pi, \cdots, \pm(\pi L-\pi)\}$.

Physically, the factorized form of the interaction term ensures that the pairing model has a sensible form also in
real space, where it describes long-range pair hopping.  Computationally, the separable form allows for the RGK Hamiltonian to be represented exactly as a Matrix Product Operator (MPO) of bond dimension 4 and thus to be efficiently solved with DMRG even when the exact solvability condition $4\eta_k^2=\epsilon_k+2t_1+2t_2$ is not fulfilled.
 
If instead the latter condition is realized, the RGK Hamiltonian assumes the standard form of a hyperbolic Richardson-Gaudin integrable model \cite{Dukelsky2004Aug, Ortiz2005Feb}. Its eigenstates involving $N_p$ fermion pairs and $N_\nu=0$ or 1 unpaired fermions are then given by
\begin{equation}
\label{bethe_RGK}
\left|\psi\right\rangle=\prod_{\alpha=1}^{N_p}\left(\sum_{k \in \mathcal{S}_{k+}^\phi} \frac{\eta_k}{\eta_k^2-e_\alpha} {c}_k^{\dagger} {c}_{-k}^{\dagger}\right)|\nu\rangle \quad \text{(any }G)
\end{equation}
where, similar to Eq. (\ref{richardson}), the spectral parameters $e_\alpha$ are determined by solving a set of nonlinear equations.

The criterion introduced in Ref. \cite{Ortiz2014Dec} for establishing the emergence of topological superfluidity in many-fermion particle-number conserving systems uses the ground state energies for $N$ and  $N\pm1$  particles (for both periodic
and antiperiodic boundary conditions) to identify the relevant fermion parity switches. For a superconductor to be topologically nontrivial, the ground state energies for even and odd number-parity need to cross an odd number of times when increasing the enclosed flux (in ring geometry) from $\phi=0$ to $\phi=2\pi$ (in units of $\Phi_0 = h/2e$). Actually, the information on the ground state fermion parity for only these two values of the flux is sufficient to determine whether the number of crossings is even or odd.

More concretely, to identify the fermion parity switches one computes the ground state energy $E_{\text{gs}}^{\phi}(N)$ for a number $N$ of fermions in the chain of $L$ sites, imposing periodic ($\phi=0$) or antiperiodic ($\phi=2\pi$) boundary conditions, and compares
\begin{equation}
\label{evenodd}
\begin{aligned}
E_{\text{odd}}^{\phi}&=[E_{\text{gs}}^{\phi}(N+1)+E_{\text{gs}}^{\phi}(N-1)]/2~,~\text{and}\\ 
E_{\text{even}}^{\phi}&=E_{\text{gs}}^{\phi}(N)~,
\end{aligned}
\end{equation} assuming $N$ is even. Their difference determines the fermion parity, which has opposite sign at $\phi=0$ and $\phi=2\pi$ in the topologically nontrivial phase.

\subsection{RGK chain: RBM results}

 \begin{figure}[ht!]
\centering
\includegraphics[width=0.5\textwidth]{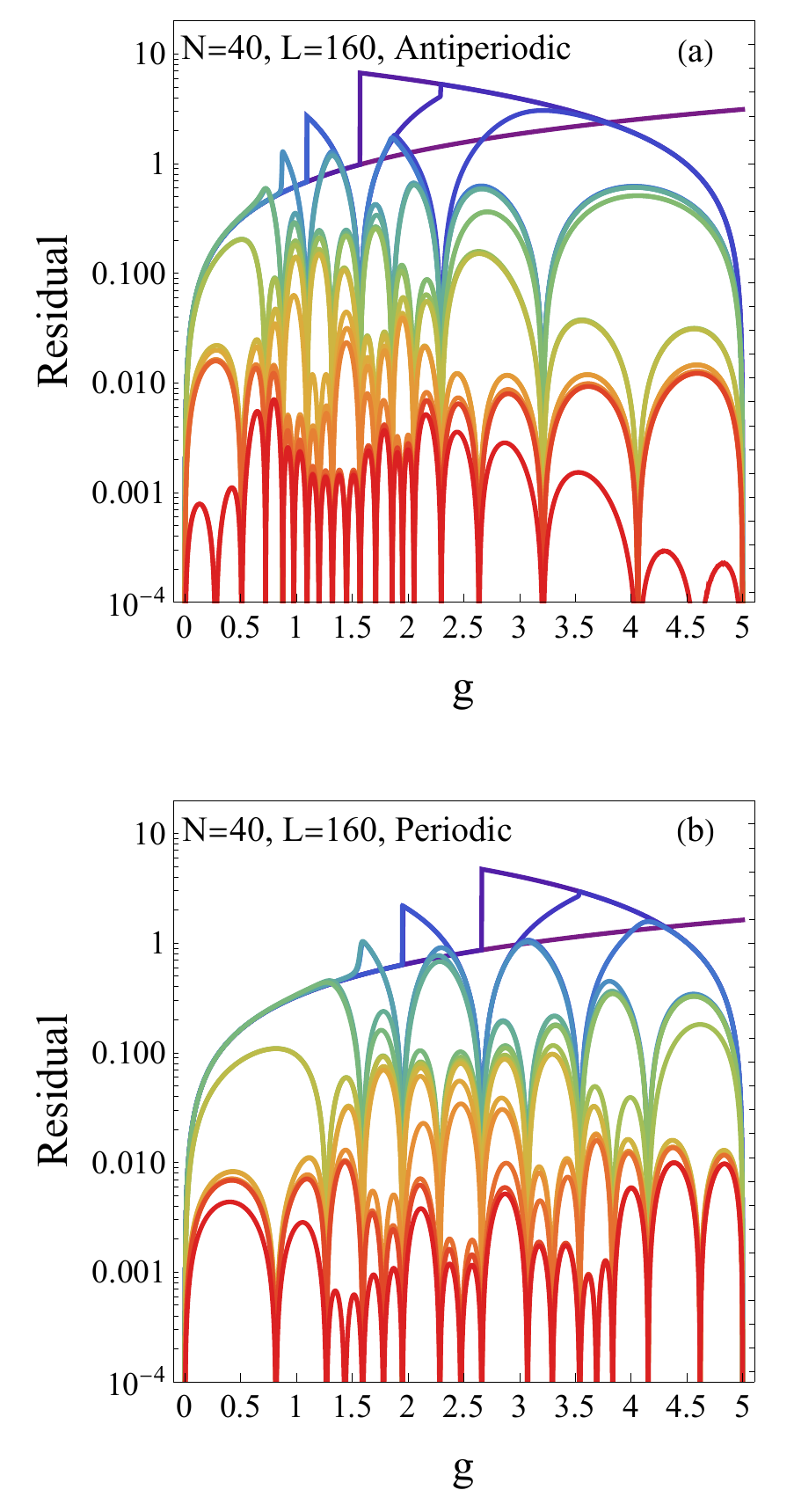}
\caption{Evolution of the Residual (\ref{res}), in units of $t_1$, during the offline emulation phase of the RGK Hamiltonian (\ref{RKG}) for a chain of $L=160$ sites at quarter filling with antiperiodic (top panel) and periodic boundary conditions (bottom panel).}
\label{fig8}
\end{figure} 

 \begin{figure}[ht!]
\centering
\includegraphics[width=0.5\textwidth]{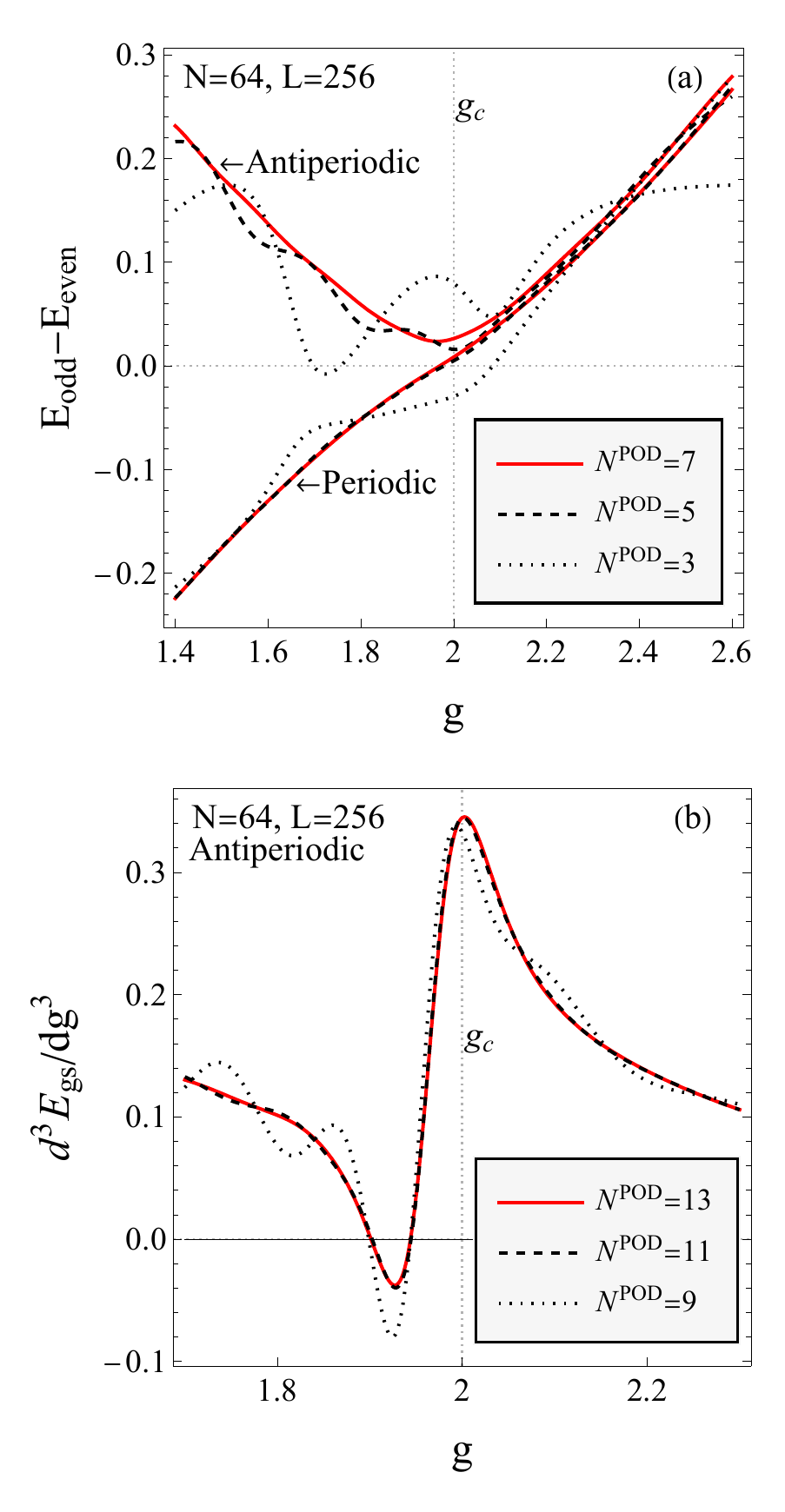}
\caption{Even-odd energy differences (\ref{evenodd}), top panel, and ground state energy third derivative, bottom panel, in units of $t_1$, for various reduced bases of dimensionality $N^{\text{POD}}$ obtained upon POD compression for a RGK chain of $L=256$ sites at quarter filling. The thermodynamic limit critical strength $g_c=2$ is evidenced.}
\label{fig9}
\end{figure}

 \begin{figure}[ht!]
\centering
\includegraphics[width=0.5\textwidth]{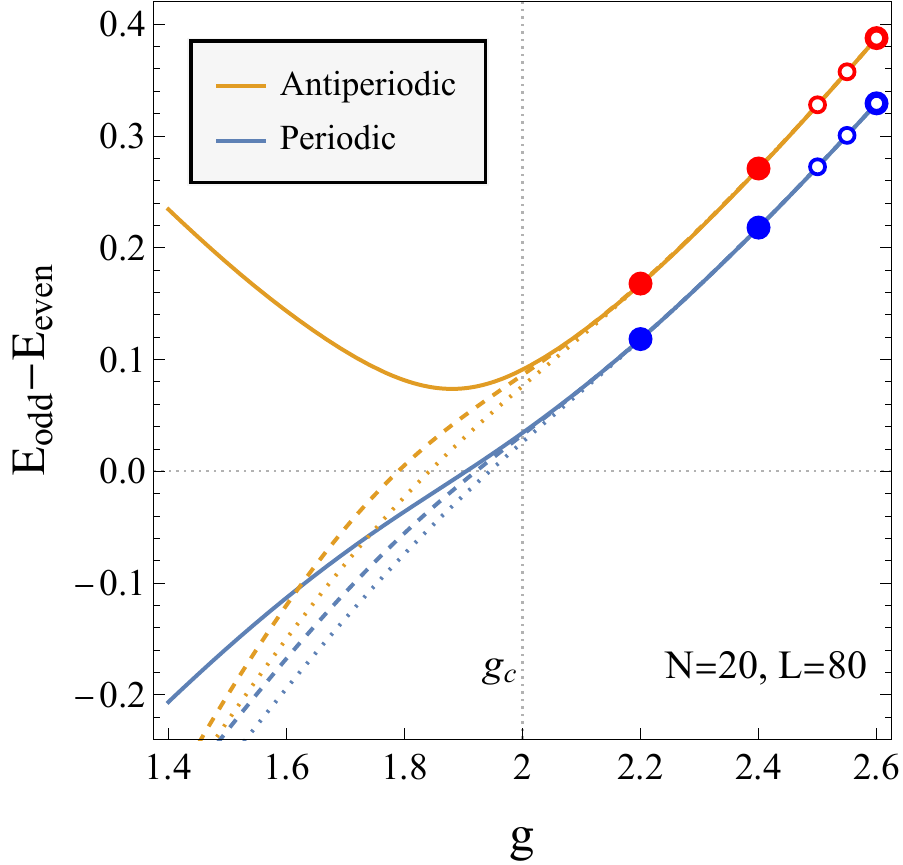}
\caption{Extrapolated  even-odd energy difference (\ref{evenodd}), in units of $t_1$, for a RGK chain on $L=80$ sites at quarter filling (dotted curve for open circle sampling points and dashed curve for filled dot sampling points). The thermodynamic limit critical strength $g_c=2$ is evidenced.}
\label{fig10}
\end{figure}

To benchmark the RBM capabilities in emulating the RGK chain across its various phases, we follow Ref. \cite{Ortiz2014Dec} and take $t_1=1$ as the unit of energy  and set $t_2=0$; furthermore, we consider the chain at quarter filling $N/L=1/4$. For this specific value, the critical reduced strength 
\begin{equation}
g=GL/2    
\end{equation}
corresponding to the third-order topological phase transition (in the thermodynamic limit) is $g_c=2$.

We present in Fig. (\ref{fig8}) a typical evolution of the Residual profile during the sampling of the interval $0\leq g \leq5$ using the greedy algorithm described in Section \ref{s2}, with an emulation accuracy threshold of $0.01$. We observe how the ground state evolution with increasing $g$ is strongly dependent on the boundary conditions: in the antiperiodic case the range of greatest variation (with the highest density of sampling points) is observed to be $1<g<2$ preceding the phase transition, while in the periodic case there is a pronounced variation on both sides of the phase transition point, more or less within $1<g<4$. 

In the following, we restrict our attention to the interval $1.4<g<2.6$ and show in Fig. (\ref{fig9}) the signatures of (the precursor to) the topological phase transition in a system of $N=64$ particles on $L=256$ sites. The RBM emulator is able to account well for both the gross features of the phase transition (top panel) and for the finer details such as correctly capturing the order of the phase transition (lower panel, note that in the case of a finite system the discontinuity in the third-order derivative is naturally smoothed out). 

In order to reach the 0.01  emulation accuracy threshold (as given by the Residual's maximum) for each case with $N=63$, 64, 65 and periodic/antiperiodic boundary conditions necessary to obtain Fig. (\ref{fig9}), no more than 13 sampling points are required in the interval $1.4<g<2.6$ during the offline step. In Fig. (\ref{fig9}) we also confirm that the gross features of the phase transition are still present when performing a moderate POD compression (of the final reduced basis obtained in the offline step), while the more sensitive energy third derivative requires an almost lossless description in order to be faithfully reproduced.

In concluding this section, we remark that an emulator trained only within the topologically trivial phase cannot extrapolate well into the topologically nontrivial phase even for a relatively small system, as displayed in Fig. (\ref{fig10}). This is contrary to one of the earliest applications of Eigenvector Continuation in Ref. \cite{frame2018} where the extrapolation from the dilute gas regime of a 3D Hubbard model was possible well into the clusterized regime (for a small four-particle system). In our specific topological transition, the failure of the extrapolation may be diagnosed by the large concentration of sampling points in Fig. (\ref{fig9}), especially for the case of antiperiodic boundary conditions, indicating a rapid and significant change in the ground state structure in the phase transition region. 

We leave a more detailed study of the RGK chain, within the DMRG+RBM approach, away from exact solvability and in various $t_2/t_1$ regimes to a future work, and investigate in the next section another variation on the Richardson pairing model particularly relevant to the state-of-the-art technological applications of condensed matter phenomena.

\section{emulating Super-Semi Hybrid Devices}
\label{s5}

\subsection{Mesoscopic superconductor-semiconductor hybrid devices}

As a second generalization of the Richardson Hamiltonian, we consider in this section its coupling to an impurity
level: this introduces pair-breaking processes and breaks integrability, but nevertheless the resulting model may still be accurately solved using DMRG \cite{Pavesic2021}. The physical situation  corresponds to a Quantum Dot (QD) coupled to a small superconducting island (SIs) into a strongly-correlated hybrid device where the three-way competition of pairing correlations, Coulomb blockade and Kondo screening results in a richness of emergent phenomena (see e.g. Refs. \cite{Fang2022Aug, Pavesic2021} and references therein). In particular, for a small mesoscopic superconducting island with a considerable charging energy and strong even-odd occupancy effects, the subgap states have properties quite unlike those of the standard Yu-Shiba-Rusinov (YSR) states \cite{EstradaSaldana2022Apr} formed inside a macroscopic superconductor's spectral gap by binding a Bogoliubov quasi-particle at the impurity \cite{yu1965bound, shiba1968classical,rusinov1969superconductivity, sakurai1970comments}. \\

While the theoretical results are in excellent qualitative agreement with experiment regarding the many-body properties of such devices \cite{Saldana2022Feb,Saldana2022Apr,Malinowski2022Oct}, their efficient modelling is limited by the large number of independent parameters, i.e. SI and QD charging energies and gate-defined occupancies,  QD-SI tunnel and capacitive coupling strengths, QD and SI effective $g$-factors or spin-orbit coupling strengths. With the only studies currently available having been carried out for the simplest QD-SI and SI-QD-SI configurations, the efficient modelling of more realistic systems involving multiple interconnected SIs and QDs could be enabled by the RBM-based emulation. 

As a proof-of-concept application of the RBM to accelerate the theoretical modelling of this class of hybrid devices, we consider in this section the QD-SI system of Refs. \cite{Pavesic2021, EstradaSaldana2022Apr}. The total Hamiltonian is obtained by combining the Anderson model for the interacting quantum dot and the Richardson model for the small superconducting island into
\begin{equation}
    \label{QDSIHam}H=H_{\mathrm{QD}}+H_{\mathrm{SI}}+H_{\mathrm{hyb}}+H_{\mathrm{cc}}
\end{equation}
where the QD, SI, hybridization and capacitive coupling terms are
\begin{equation}
\begin{aligned}
H_{\mathrm{QD}} &=\epsilon_{\text{QD}}\,  {N}_{\mathrm{QD}}+U \,{N}_{\mathrm{QD}, \uparrow} \,{N}_{\mathrm{QD}, \downarrow} \\
&=(U / 2)\left({N}_{\mathrm{QD}}-\nu\right)^2+\mathrm{const}, \\
H_{\mathrm{SI}} &=H_{\text{Richardson}}+E_c\left({N}_{\mathrm{SI}}-n_0\right)^2, \\
H_{\mathrm{hyb}} &=\frac{t}{\sqrt{L}} \sum_{i=1}^L \sum_{\sigma=\uparrow,\downarrow}\left(c_{i \sigma}^{\dagger} d_\sigma+\text {H.c.}\right)~,\\
H_{\mathrm{cc}} &=U_{\text{cc}}\left({N}_{\mathrm{QD}}-\nu\right)\left({N}_{\mathrm{SI}}-n_0\right)~.
\end{aligned}
\end{equation}
For the quantum dot,  $N_{\text{QD},\sigma}=d^\dagger_\sigma d_\sigma$ is the number operator for each spin projection $\sigma=\uparrow,\downarrow$, and $N_{\text{QD}}=N_{\text{QD},\uparrow}+N_{\text{QD},\downarrow}$. The gate-defined optimal QD occupation $\nu$ is related to its single particle energy $\epsilon_{\text{QD}}$ and electron-electron repulsion strength $U$ by $\nu=1/2-\epsilon_{\text{QD}}/U$. The SI pairing correlations are modelled by the Richardson Hamiltonian of Eq. (\ref{ham1}) which involves here $L=50$ doubly degenerate levels spaced by $d=2D/L$, where $D$ is the half bandwidth (in this section we follow Ref. \cite{Pavesic2021} and take $D=1$ as the unit of energy). We tune the pairing strength to $G = 0.4d$ which gives rise to the superconducting gap
$\Delta= 0.166$ and ensures that an appropriate
number of levels participate in the pairing interaction, thus
reducing the finite-size effects \cite{Pavesic2022}.  The SI Coulomb repulsion effects are taken into account by the second term in $H_{\text{SI}}$ written in terms of its charging energy $E_c$ and its gate-defined optimal occupation $n_0$. We fix the QD-SI tunneling amplitude $t$ by choosing an intermediately strong value for the hybridization strength $\Gamma=\pi \rho t^2=0.4\Delta$, where $\rho=1/2D$ is the SI normal state density of states.

\subsection{QD-SI emulation: RBM results}

 \begin{figure}[ht!]
\centering
\includegraphics[width=0.5\textwidth]{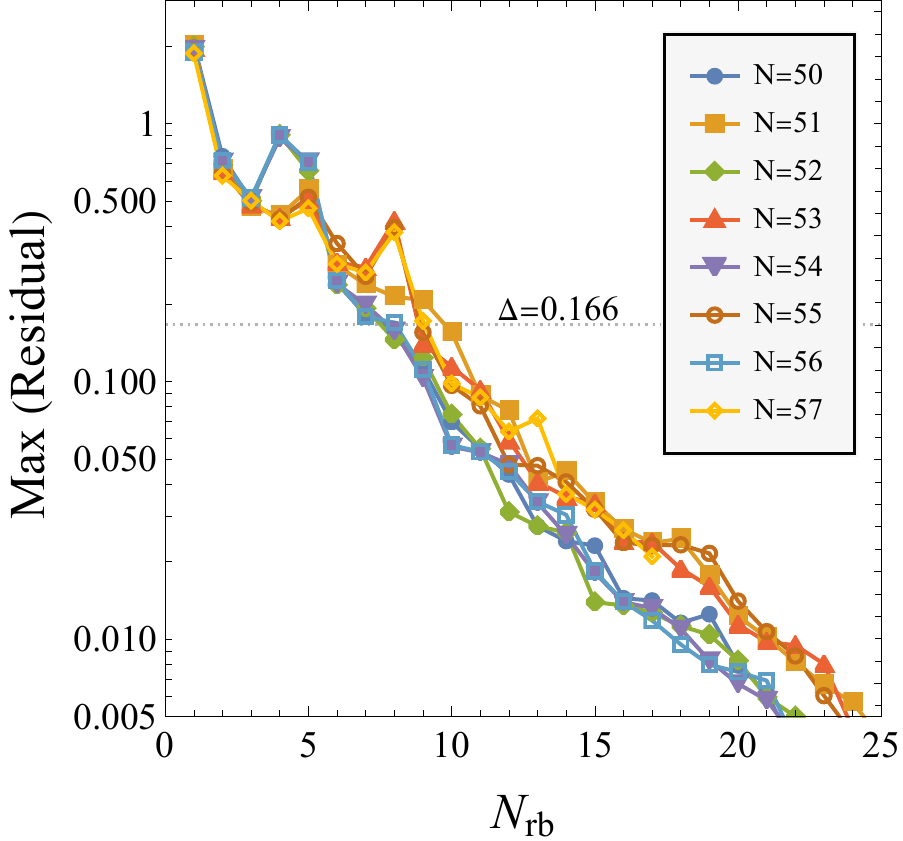}
\caption{Evolution of the Residual's maximum, in units of $D$, versus the reduced basis dimension $N_{\text{rb}}$ during the offline emulation phase of the QD-SI Hamiltonian (\ref{QDSIHam}) for various particle numbers $N$.}
\label{fig11}
\end{figure} 

 \begin{figure*}[ht!]
\centering
\includegraphics[width=\textwidth]{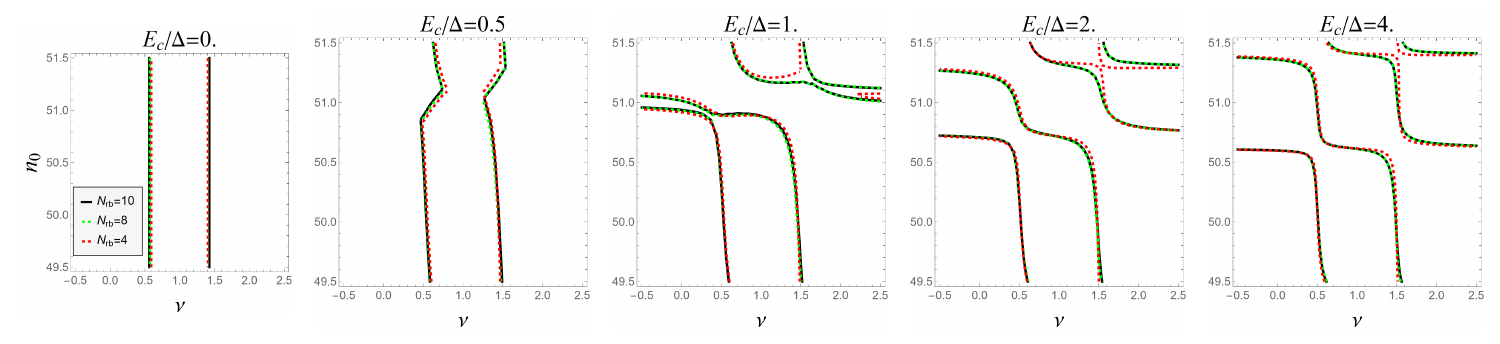}
\caption{QD-SI ground state parity transition curves as a function of gate voltages applied to the QD ($\nu$) 
and to the SI ($n_0$) for varying SI charging energy $E_c$ and for different reduced basis dimensionalities $N_{\text{rb}}$. We consider fixed $U=4\Delta,~U_{cc}=0$.}
\label{fig12}
\end{figure*} 

 \begin{figure*}[ht!]
\centering
\includegraphics[width=\textwidth]{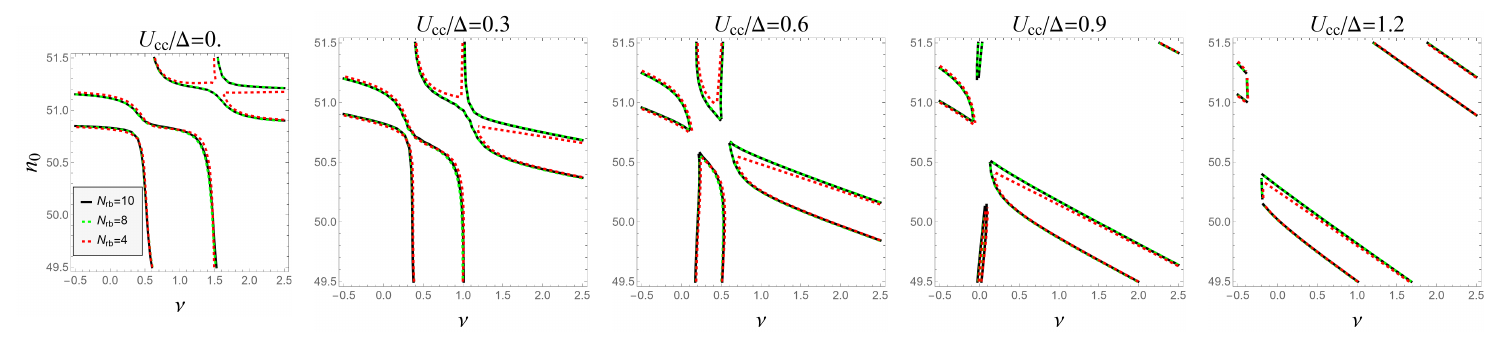}
\caption{QD-SI ground state parity transition curves as a function of gate voltages applied to the QD ($\nu$) 
and to the SI ($n_0$) for varying QD-SI capacitive coupling strength $U_{cc}$ and for different reduced basis dimensionalities $N_{\text{rb}}$. We consider fixed $U=4\Delta,~E_c=1.3\Delta$.}
\label{fig13}
\end{figure*} 

In this application, we take the control parameters for the RBM emulation to be $U,\nu, E_c, n_0$ and $U_{\text{cc}}$.  By using the particle number conservation $[H,N]=0$ with $N=N_{\text{QD}}+N_{\text{SI}}$, we may restrict our attention  to each particle number sector individually and absorb the effects of the SI's charging energy into $H_{\text{QD}}$ by a redefinition the QD parameters. We thus work with the effective Hamiltonian
\begin{equation}
\begin{aligned}
    H(\xi_1,\xi_2)&=H_0+\xi_1N_{\text{QD}}+\xi_2 N_{\text{QD}}^2~,
    %&=H_0+\overline{U}\left(N_{\text{QD}}-\overline{\nu}\right)^2+\text{const.}
    \end{aligned}
\end{equation}
where $H_0=H_{\text{Richardson}}+H_{\text{hyb}}$ and
\begin{equation}
    \begin{aligned}
    \xi_1&=(U_{\text{cc}}-U)\,\nu - (2E_c+U_{\text{cc}}) (N-n_0)~,\\
    \xi_2&=U/2+E_c-U_{\text{cc}}~,
    %\overline\nu&=-\xi_1/(2\xi_2)~,~\overline{U}=\xi_2~,
    \end{aligned}
\end{equation}
reducing the number of independent control parameters to two. We restrict our analysis to the parameter domain $(\xi_1,\xi_2)\in [-5.8,11.6]\times [0,2.5]$  which is sufficient to cover the physical domain $(n_0,\nu)\in [49.5,51.5]\times [-0.5,2.5]$  up to moderately large values of $U,E_c$ and $U_{cc}$ for all relevant particle numbers $N\in\{50,51,...,57\}$.\\
 
We show in Fig. (\ref{fig11}) the Residual's maximum exponential decrease during the greedy learning algorithm which confirms the efficient RBM representation of the QD-SI parameter space. The learning rates are almost identical for the range of system sizes considered, which is consistent with the reduced finite size effects. Additionally, we observe that the emulation for odd-parity states requires a slightly larger number of sample points than for even-parity states in order to reach the same accuracy (as quantified by the Residual's maximum).

Remarkably, a very small basis dimensionality (corresponding to a modest emulation accuracy of the order of the superconducting gap) is sufficient for reaching convergence in emulating the QD-SI charge stability diagram, as seen in Figs. (\ref{fig12}) and (\ref{fig13}). The evolution of the ground state parity transition lines (by ``ground
state'' we mean the lowest energy state over all particle
number sectors) is shown in Fig. (\ref{fig12}) for varying SI charging energy $E_c$, and in Fig. (\ref{fig13}) for varying capacitive coupling strength $U_{cc}$. In the former case, our emulated results agree with those presented in Fig. 1 of Ref. \cite{Pavesic2021}, regarding the evolution from the YSR regime (with $2e$ periodicity along
the $n_0$ axis) to the Coulomb blockaded regime (with $1e$ periodicity). In the latter case, the phase boundaries acquire an angle due to the capacitive coupling influencing the occupancy in both parts of the system, in agreement with the findings in Refs. \cite{EstradaSaldana2020Nov, EstradaSaldana2022Apr}.\\

\begin{figure}[ht!]
\centering
\includegraphics[width=0.5\textwidth]{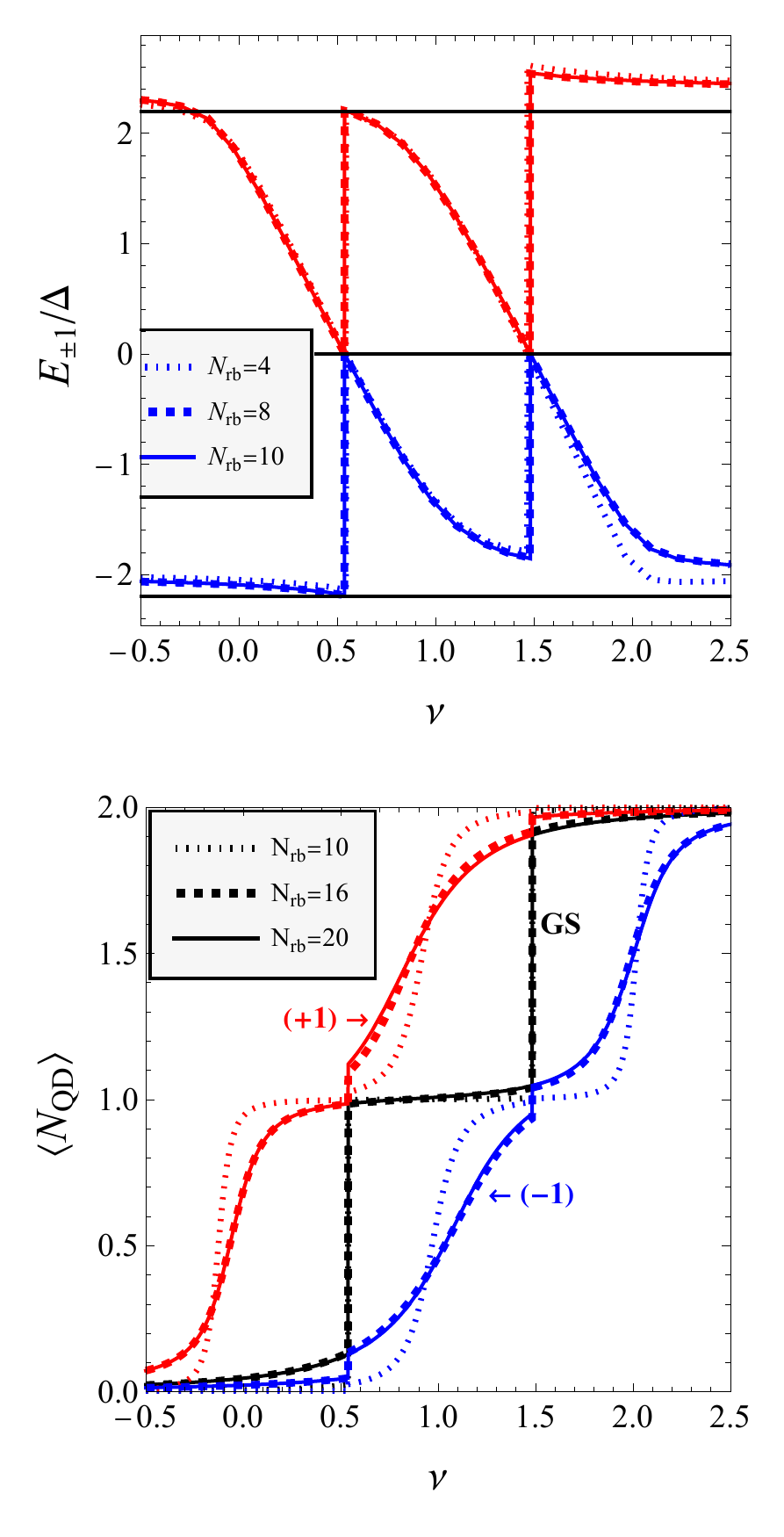}
\caption{Excitation energies $E_{\pm1}$ (top panel) and quantum dot average occupations $\langle N_{\text{QD}}\rangle$ (bottom panel) as a function of the gate voltage applied to the QD ($\nu$) for the ground state (GS) and the lowest-energy (subgap)
excited particle-like and hole-like states $(\pm 1)$, for even $n_0=52$ and for different reduced basis dimensionalities $N_{\text{rb}}$. We consider fixed $U=4\Delta,~E_c=1.2\Delta$. The solid horizontal black lines in the top panel are situated at 0 and $\pm( 1+E_c/\Delta)$.}
\label{fig14}
\end{figure}

The occupancies constitute a more sensitive metric for the convergence of the reduced basis emulation than the energy itself. The $\nu$ dependence of the emulated average quantum dot occupancy (for even $n_0$) is shown in Fig. (\ref{fig14}) to be in  agreement with Fig. (5a) of Ref. \cite{Pavesic2021}. As expected, compared to the energy emulation, a reduced basis of a slightly increased dimensionality is necessary to achieve convergence in the RBM description of the occupancy. This corresponds to about an order of magnitude lower residual maximum of about $0.1\Delta$, as can be read from Fig. (\ref{fig11}). Nevertheless, the RBM curves accurately capture the rapid variation of the lowest-energy (subgap)
excited states around half filling, which is directly related to an
equally rapid variation of the excitation spectrum \cite{Pavesic2021}, as also shown in Fig. (\ref{fig14}). The latter agrees qualitatively with  Fig. (4d) of Ref. \cite{Pavesic2021}, with the only quantitative difference being that our excitation energies deviate slightly from the values $\pm(1+E_c/\Delta)$ at small/large values of $\nu$, due to finite size effects.

\section{Conclusions}
\label{s6}

This work provides the first indications on the RBM capability to capture efficiently the ground state correlations induced by the pairing force across entire parameter spaces of various interacting many-body systems. 

Very small bases were confirmed to accurately describe the weak-to-strong pairing cross-over in the Richardson model (in Section \ref{s3}), the third-order topological phase transition of the interacting Richardson-Kitaev chain (in Section \ref{s4}, and
the charge stability diagram of a hybrid quantum dot - superconductor device (in Section \ref{s5}). In all cases the efficient sampling of the relevant parameter space was confirmed by the exponential decay of the Residual (\ref{res}) during the construction of the reduced basis. 

On the one hand, the Residual is a direct measure for the emulation error; its magnitude is given by a combination of the error in the energy and the error in the ground state eigenvector itself. Thus, it is typically larger than each of the two individual errors, which in practice ensures that the actual energy error will be much smaller than the chosen Residual threshold for the reduced basis construction, e.g. compare Figs. (\ref{fig1}) and (\ref{fig6}b) and see also Ref. \cite{sarkar2022}.  

On the other hand, the distribution of sampling points generated by the greedy Residual optimization strategy may provide qualitative insight into the evolution of the ground state structure across the parameter domain. A higher density of sampling points implies a rapid change in the eigenstate structure, specific to cross-over phenomena, see e.g. the discussion around Fig. (\ref{fig1}), or quantum phase transitions, see e.g. the discussion around Fig. (\ref{fig8}) and also Ref. \cite{rizzi2022}. 

More generally, beyond investigating phase diagrams as outlined in Ref. \cite{rizzi2022}, the RBM could actually enable the otherwise expensive but desirable microscopic models to better interface with experimental design and data analysis. For example, the precise location of possible qu$d$it operational sweet-spots (regions with reduced noise sensitivity; see also \cite{Pavesic2022}) for more realistic systems involving multiple interconnected SIs and QDs could be precisely determined by the fast and accurate RBM scans of their enlarged parameter spaces (impractical only with direct DMRG solvers). Conversely, fitting the model parameters to the experimental data, as in Refs. \cite{EstradaSaldana2022Apr,Saldana2022Feb} for this class of systems, could become effortless once a cheap RBM effective representation is found. 

Perhaps a more challenging novel RBM application would consist of building emulators able to accelerate transport simulations (e.g. those of Ref. \cite{Chung2022Sep}) upon adapting the RBM treatment of time-dependent differential equations \cite{Hesthaven2022May}. This would be a valuable tool in mitigating the rise in computational complexity when extending the transport simulation to finite temperature, where more demanding density matrix or thermofield descriptions are necessary (see  \cite{Chung2022Sep} and references therein).

\begin{acknowledgments}
We warmly thank Jens Paaske, Max Geier, Michele Burrello, Matteo Rizzi, Andrea Maiani, Ben Joecker, Luka Pave{\ifmmode\check{s}\else\v{s}\fi}i{\ifmmode\acute{c}\else\'{c}\fi}, Rok  {\ifmmode\check{Z}\else\v{Z}\fi}itko, Jorge Dukelsky and Alexandru Nemnes for useful discussions and suggestions. V.V.B. acknowledges many fruitful research interactions at the Condensed Matter Group and QDev (Niels Bohr Institute, Copenhagen University). This work was supported by the grant of the Romanian Ministry of Education and Research, CNCS - UEFISCDI,
project number  PN-III-P4-ID-PCE-2020-1142, within PNCDI III.
\end{acknowledgments}

\appendix

\section{}
\label{app}
In this section we consider a dynamic renormalization of the Residual designed to assess the emulation accuracy on the relevant energy scale  for each region of the parameter space.

Due to the Residual defined in Eq. (\ref{res}) scaling as an energy, the greedy algorithm based on it may preferentially sample the strong-coupling regime (large $|f_p(\boldsymbol{\xi})|$ in eq. \ref{aff}) in order to reach the imposed global energy-based threshold for the emulation accuracy. This over-sampling may be observed for instance in Figs. \ref{fig6} and \ref{fig7}, where the final errors of the correlation energy and of the canonical gap (relative to their exact values) are seen to decay exponentially into the strong-pairing (large-$G$) regime of the Richardson pairing Hamiltonian (\ref{ham1}), for a $0.01\epsilon$ global emulation accuracy threshold.

We could allow instead for the emulation accuracy to be measured locally against the relevant energy scale for each region of the parameter space, thus imposing a global threshold for the relative emulation accuracy. This could avoid the previously mentioned over-sampling issue, thus potentially providing an even more efficient reduced basis construction methodology.

To illustrate these concepts with a specific example, we consider once again the Richardson pairing Hamiltonian of Eq. (\ref{ham1}) with the pairing strength $G$ as the control parameter, $\xi=G$.  As the renormalized version of the Residual, we work in this Appendix with the definition
    \beq
    \label{res2}
    \text{Res}(G)\equiv\frac{\left\lVert\mathcal{H}(G)|\psi^{(\text{rb})}\rangle-E^{(\text{rb})}|\psi^{(\text{rb})}\rangle\right\rVert}{|E^{(G=0)}-E^{(\text{rb})}|}~,
    \eeq
with all reduced-basis (rb) quantities having an implicit $G$-dependence. As a natural local energy scale, we are thus considering the gain in energy obtained when turning on the pairing interaction, $E^{(G=0)}-E(G)$. This still needs to be evaluated in the reduced-basis approximation for the error estimation procedure to remain efficient.

\begin{figure}[ht!]
\centering
\includegraphics[width=0.5\textwidth]{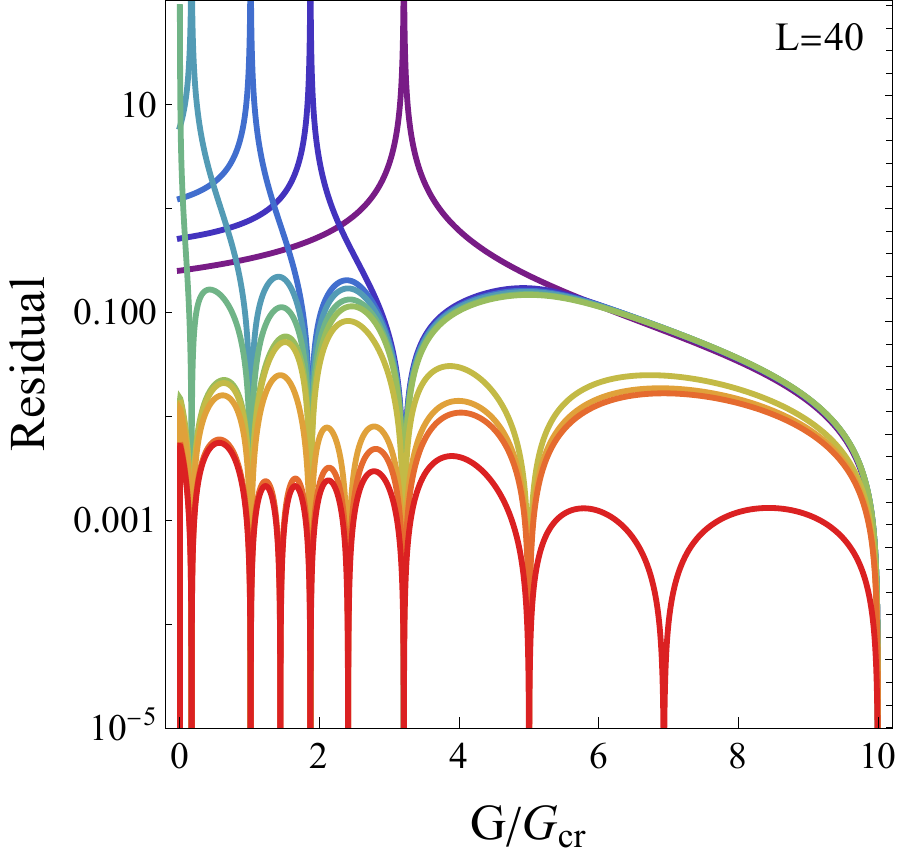}
\caption{Evolution of the Residual (\ref{res2}) during the offline emulation phase of the pairing Hamiltonian (\ref{ham1}) for a system of $N_p=20$ pairs distributed over $L=40$ levels. The color spectrum (violet, blue, green, orange, red) correlates with the increasing iteration number of the greedy self-learning algorithm.}
\label{fig15}
\end{figure}

The evolution of the renormalized Residual profile during the greedy sampling algorithm is shown in Fig. \ref{fig15}, for the particular choice of $G=10G_{\text{cr}}$ as a first  sampling point (the $G=0$ Hartree-Fock state leads to an inconclusive constant renormalized residual at the first iteration). As seen in Fig. \ref{fig15}, our chosen 0.01 relative emulation accuracy threshold is achieved with only 10 sampling points. During the first few iterations (until the $G=0$ HF state is included in the basis), new sampling points are selected by the vanishing of the denominator in Eq. (\ref{res2}). This signals  where the reduced-basis energy approximation becomes worse than its uncorrelated value, thus prompting for a new exact evaluation.

\begin{figure}[ht!]
\centering
\includegraphics[width=0.5\textwidth]{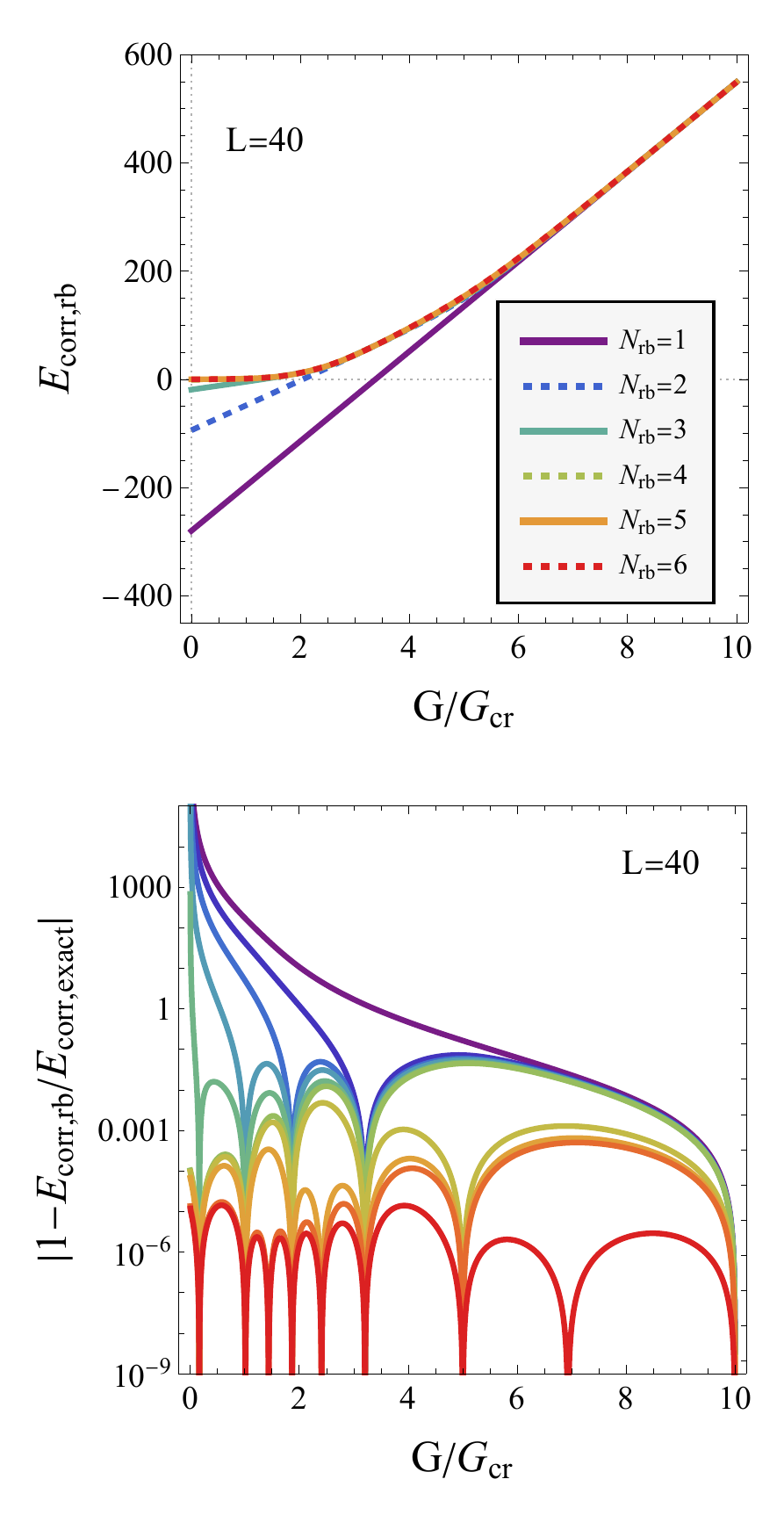}
\caption{Evolution of the reduced basis correlation energy $E_\text{corr}$ (\ref{ecorr}) (top panel), in units of the level spacing $\epsilon$, and the corresponding error relative to its exact value (bottom panel) during the offline emulation phase.}
\label{fig16}
\end{figure}

Correspondingly, in Fig. \ref{fig16} the correlation energy of Eq. (\ref{ecorr}), $ E_{\text{corr}}(G)\equiv  E_{\text{HF}}(G)-E(G)$, is shown to become negative during the first iterations, with its zeroes closely related to the residual's peaks in Fig. \ref{fig15}. Due to the vanishing of the exact correlation energy at $G=0$, the corresponding relative error diverges here until the Hartree-Fock state is included in the basis (which is needed to ensure a sufficiently accurate description of the weak pairing regime).

\begin{figure}[ht!]
\centering
\includegraphics[width=0.5\textwidth]{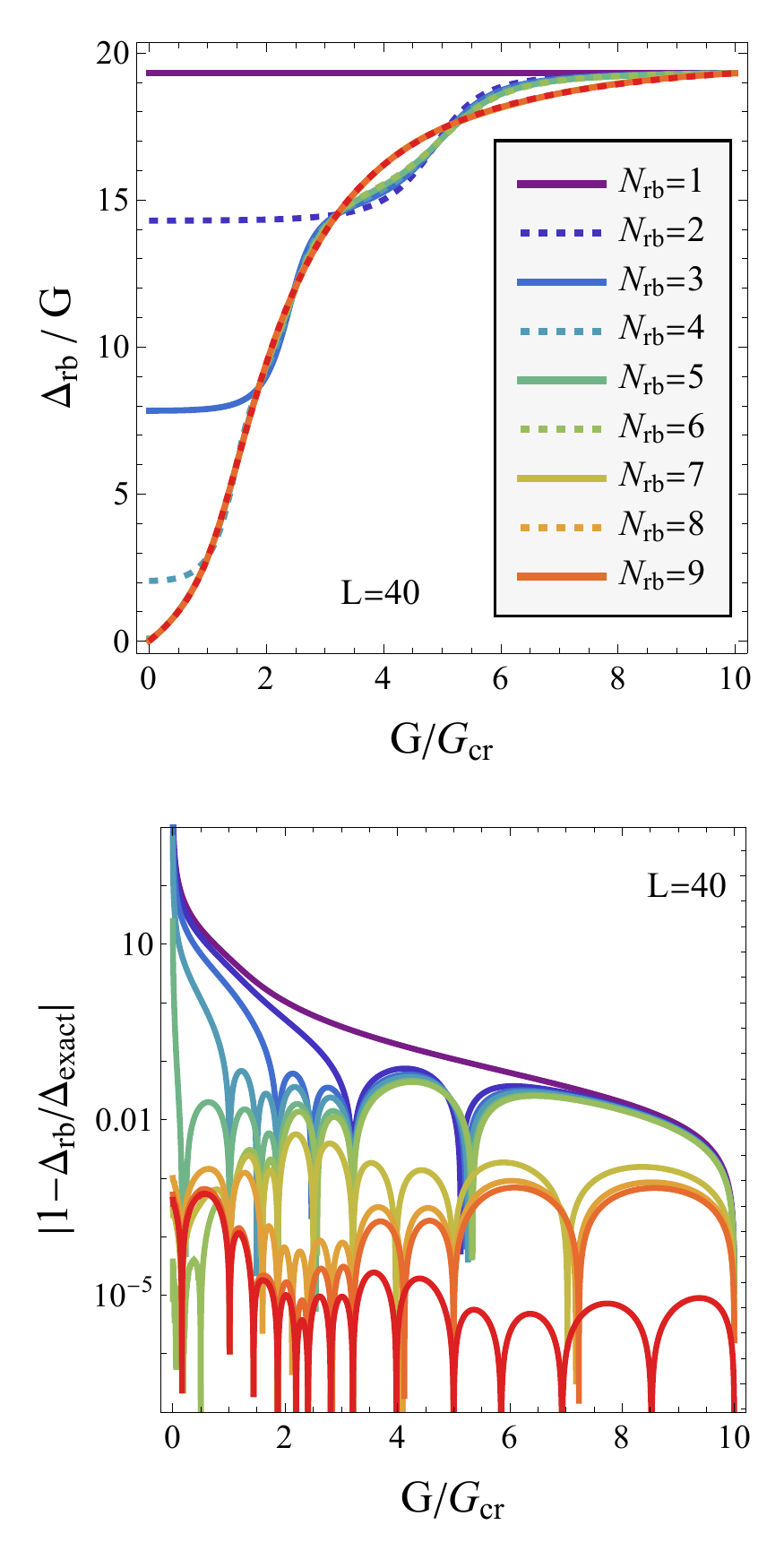}
\caption{Evolution of the reduced basis canonical gap $\Delta$ (\ref{gap}), (top panel), and the corresponding error relative to its exact value, (bottom panel), during the offline emulation phase.}
\label{fig17}
\end{figure}

Remarkably, for both the correlation energy and for the canonical gap of Eq. (\ref{gap}), the local maxima in the final relative errors exhibit only only small variations across the entire $G$-interval, as seen in Figs. \ref{fig16} and \ref{fig17} (compare with Figs. \ref{fig6} and \ref{fig7} obtained using the un-normalized residual of Eq. (\ref{res})). This confirms that the definition (\ref{res2}) is able to address the previously discussed strong-coupling over-sampling issue and thus to provide a sensible emulation accuracy measure for more efficient sampling strategies.

We finally note that the choice of Eq. (\ref{res2}) for the Residual normalization is not unique. We obtained a qualitatively similar residual profile evolution with that of Fig. \ref{fig15} (same sampling points, with smoothed residual peaks during the first iterations) by adapting the prescription of Ref. \cite{sarkar2022} and using the normalization factor $\langle \psi^{(\text{rb})}| [E^{(G=0)}-\mathcal{H}]^2|\psi^{(\text{rb})}\rangle^{1/2}$. 

\vspace{3cm}

\bibliography{mybibfile}

\end{document}